\documentclass[prl,twocolumn,showpacs,preprintnumbers,amsmath,amsfonts,amssymb,floatfix,aps,superscriptaddress]{revtex4-2}
\usepackage{graphicx}
\usepackage{xr}
\usepackage{enumitem}
\usepackage{amssymb}
\usepackage{amsmath}
\usepackage{amsfonts}
\usepackage{bm}
\usepackage{dsfont}
\usepackage{comment}
\usepackage{color}
\usepackage{relsize}
\usepackage{float}
\usepackage{bm}
\usepackage{braket}
\usepackage[normalem]{ulem}
\usepackage[nodisplayskipstretch]{setspace}
\usepackage{tikz}
\usepackage{hyperref}
\hypersetup{
     colorlinks=true,
     linkcolor=blue,
     filecolor=blue,
     citecolor = blue,
     urlcolor=blue,
   }
\usepackage[export]{adjustbox}
\usepackage{units}
\usepackage[left]{lineno}
\usepackage{xspace}

\newcommand{\su}[1]{\ensuremath{\mathrm{SU}(#1)}\xspace}
\newcommand{\sun}{\su{N}}
\allowdisplaybreaks
\begin{document}
	\title{Engineering SU($N$)-Symmetric Hubbard Models with Microwave-Shielded \\ Dipolar Molecules}

\author{Jing-Lun Li}%
\email{jinglun.li@ist.ac.at}
\affiliation{Institute of Science and Technology Austria (ISTA), 
Am Campus 1, 3400 Klosterneuburg, Austria
}%

\author{Ragheed Alhyder}%
\email{ragheed.alhyder@ist.ac.at}
\affiliation{Institute of Science and Technology Austria (ISTA), 
Am Campus 1, 3400 Klosterneuburg, Austria
}%

\author{Andreas Schindewolf}
\affiliation{Vienna Center for Quantum Science and Technology,
Atominstitut, TU Wien, Stadionallee 2, 1020 Vienna, Austria}

\author{Kaden~R.~A. Hazzard}
\affiliation{Department of Physics and Astronomy, Rice University, Houston, TX, USA
}%
\affiliation{Smalley-Curl Institute, Rice University, Houston, TX 77005, USA
}%

\author{Mikhail Lemeshko}
\affiliation{Institute of Science and Technology Austria (ISTA), 
Am Campus 1, 3400 Klosterneuburg, Austria
}%

\author{Georgios M. Koutentakis}
\email{georgios.koutentakis@ist.ac.at}
\affiliation{Institute of Science and Technology Austria (ISTA), 
Am Campus 1, 3400 Klosterneuburg, Austria
}%

\begin{abstract}
Ultracold polar molecules provide strong, long-range interactions that microwave shielding make tunable and nearly nuclear-spin independent, giving an emergent \sun symmetry.
However, extended Hubbard models of polar molecules in optical lattices lack, so far, controllable finite on-site interactions, a key ingredient of strong correlated physics. We show that tuning the Rabi frequency of the microwave coupling can bring two individual molecules (monomers) on neighboring lattice sites into resonance with a field-linked dimer (doublon) on one of the sites, enabling coherent doublon--monomer-pair conversion. In this model, we characterize the key Hubbard parameters and the dimer lifetime, demonstrating that the on-site and off-site interactions can be tuned nearly independently through the microwave amplitude and orientation, respectively. Our results provide a roadmap for implementing \sun-symmetric extended Hubbard models with controllable doublon fluctuations, providing access to quantum-simulation in the strongly dipolar regime.
\end{abstract}
\maketitle

Optical lattices have transformed ultracold gases into highly controllable artificial crystals \cite{block2005NP}, enabling numerous landmark studies of Hubbard-model physics \cite{Greiner2002Nature,Hart2015,Mazurenko2017,Chiu2019,Koepsell2019}.
Ultracold polar molecules extend this platform through large electric dipole moments, rich internal structures, and long-range interactions \cite{Bohn2017,Langen2024,Karman2024,Li2026}. They enable quantum simulation of strongly correlated matter \cite{Baranov2012,Wall2015}, precision tests of fundamental symmetries \cite{DeMille2024}, and quantum information processing \cite{Cornish2024}. In optical lattices, molecules can be trapped, addressed, and detected in programmable arrays \cite{anderegg2019Sc,Mortlock2026NatCommun}, while anisotropic dipolar interactions realize extended Hubbard models, dipolar magnets, and competing crystalline, supersolid, and superfluid orders \cite{Baranov2012,CapogrossoSansone2010}.

Historically, short-range loss long prevented stable, strongly interacting molecular quantum gases \cite{Ospelkaus2010,Ni2010,bause2023}. Collisional shielding overcomes this limitation by engineering a repulsive intermolecular barrier that blocks access to the lossy short-range regime. Such a barrier can be created with static electric fields~\cite{Avdeenkov2006, Wang2015,GonzalezMartinez2017,MukherjeeCaF2023,MukherjeeHutson2024}, that enabled supression of reactive losses in potassium-rubidium mixtures~\cite{matsuda_resonant_2020}, and dipolar evaporation to below the Fermi temperature in quasi-two-dimensional~\cite{valtolina_dipolar_2020}, and three-dimensional~\cite{Li2021} geometries, or with microwave dressing~\cite{Karman2018,Lassabliere2018,Anderegg2021,Buchler2007,Cooper2009,schindewolf2026review}. In the latter case, coupling a slightly blue-detuned microwave field to the lowest rotational transition dresses the molecular state inducing a lab-frame dipole moment and thereby providing control over the interaction strength~\cite{Chen2023,Li2026,schindewolf2026review}. These advances enable evaporative cooling of both fermionic~\cite{Schindewolf2022} and bosonic molecules~\cite{Bigagli2023,Lin2023}, as well as, the realization of degenerate Fermi gases~\cite{Schindewolf2022}, molecular Bose-Einstein condensates~\cite{Bigagli2024,Shi2025}, dipolar self-bound droplets~\cite{Zhang2026} and Fermi-surface deformation~\cite{Biswas2026}. 


Weak coupling between nuclear spins and dressed rotation \cite{Mukherjee2025,Will:2016,Karman:2019,Karman2025double,Deng2025,Li2026} yields nearly identical shielded potentials across nuclear-spin configurations and thus \sun symmetric interactions \cite{Mukherjee2025,Mukherjee2025R}. Compared with alkaline-earth atoms, which offer only fermionic isotopes with non-trivial \sun symmetry \cite{Congjun2003, Gorshkov2010,Cazalilla2014,IbarraGarcia2024,Polo2026}, shielded molecules offer a broader parameter space: bosonic or fermionic statistics, tunable long-range anisotropic interactions, and large internal-state manifolds \cite{Mukherjee2025,Mukherjee2025R}. In contrast, tuning interactions of alkaline-earth atoms through optical Feshbach resonances can induce dissipation and break \sun symmetry \cite{Enomoto2008,Cazalilla2014}. Loading microwave-shielded molecules into optical lattices therefore opens a route to realizing the broader class of \sun-symmetric extended Hubbard models than those available with alkaline-earth atoms~\cite{Honerkamp2004,Hermele2009,Taie2012,Taie2022,Pasqualetti2024}. Such models support rich correlated phases, including chiral spin liquids \cite{Hermele2009}, flavor-selective Mott states \cite{Chen2024}, generalized magnetic orders \cite{Taie2022}, and unconventional pairing mechanisms \cite{Rapp2007}.

Yet for molecules, lattice models still lack a key ingredient of atomic Hubbard physics: tunable on-site interactions comparable to the tunneling energy scale~\cite{baier2016}. This term controls virtual doublon-hole processes, such as superexchange \cite{Hart2015,Mazurenko2017}, pair motion, correlated hopping \cite{Schmidt2008PRL}, and soft-core density fluctuations, thereby setting magnetic energy scales and underpinning doped Hubbard \cite{Chiu2019,Koepsell2019} and \sun $t$-$J$ systems \cite{Schlomer2024PRB,Bohler2025Arxiv}. The interplay of on-site and off-site interactions can drive Wigner-Mott physics \cite{Camjayi2008NatPhys}, hidden-order insulating phases \cite{DallaTorre2006PRL}, and pair-correlated supersolids in extended Hubbard models \cite{Schmidt2008PRL,Biedron2018PRB}. In molecular systems, however, double occupation is often suppressed due to strong on-site repulsion or dynamically projected out by two-body loss, producing an effective hard-core constraint \cite{Syassen2008Science,Yan2013Observation}. While useful, this widely explored hard-core limit \cite{Goral2002,Lahaye2009,Yamamoto2012,Koziol2024}, including in dipolar 
$t-J$ models \cite{Gorshkov2011PRL,Gorshkov2011PRA,Carroll2025}, removes the doublon fluctuations needed for the broader class of Hubbard processes.

In this Letter, we show how to synthesize the missing on-site interaction with microwave-shielded molecules in an optical lattice while preserving \sun symmetry. Varying the shielding Rabi frequency and microwave detuning brings a field-linked molecular dimer \cite{fld_note} occupying one lattice site into resonance with two molecular monomers on adjacent sites. This crossing enables coherent interconversion between monomer pairs and a field-linked dimer \cite{Li2026,Chen2023,Chen:2024TM}, so that the dimer acts as an effective doublon with an on-site energy controlled by microwave parameters. 
The Rabi frequency and detuning mainly tune the on-site energy, whereas the microwave-field orientation adjusts long-range interactions, allowing nearly independent control. We derive the Hubbard parameters and dimer lifetime as functions of these control parameters, providing a guide for molecular lattice experiments.

We consider microwave-shielded fermionic polar molecules ($^{23}$Na$^{40}$K) of mass $m$ interacting via the dressed potential $V_{\text{sh}}(\bm{r})$ \cite{Li2026,Deng2023}. Under a circularly polarized microwave field in the $(xy)$ plane, the shielding potential $V_{\text{sh}}({\bm r})$ consists of a repulsive barrier at $\sim 1000\ a_0$ and an anti-dipolar long-range interaction tail that is attractive in the plane but repulsive along $z$~\cite{sm}.
It is tuned by the microwave Rabi frequency
$\Omega$ and detuning $\Delta$. Increasing $\Omega$ at fixed $\Delta/\Omega$ shifts the barrier inward while preserving its long-range interaction tail, thus deepening the in-plane attractive well, which can support bound states. We fix $\Delta/\Omega=0$ in this work.
The molecules are confined in a quasi-two-dimensional optical lattice
$V_{\mathrm{latt}}(\bm{r}) = \sum_{\mu \in \{x,y,z\}} V_{0,\mu} \sin^2(\pi r_\mu/a_{\rm latt})$,
with $V_{0,x} = V_{0,y} = V_0$, $V_{0,z} = \lambda^2 V_0$, and $\lambda > 1$ \cite{sm}. Here $a_{\rm latt}$ denotes the lattice spacing.
The microscopic many-body Hamiltonian is $\hat{H} = \hat{H}_0 + \hat{H}_{\mathrm{int}}$, with
$
\hat{H}_0 = \sum_{i=1}^{N_M} \left[ -\frac{\hbar^2}{2m} \nabla_i^2 + V_{\mathrm{latt}}(\bm{r}_i) \right]
$
and
$
\hat{H}_{\mathrm{int}} = \sum_{i<j} V_{\text{sh}}(\bm{r}_j- \bm{r}_i),
$
where $N_m$ is the number of molecular monomers.
In deep lattices, the single-particle low-energy sector is spanned by Wannier states $w_{i}(\bm{r})=w(\bm{r} - \bm{R}_i)$ localized at each site $\bm{R}_i$, with a single-particle band gap $\approx \hbar\omega_{\mathrm{ho}} = 2\sqrt{s}\,E_R$, where $s = V_0/E_R$ is the dimensionless in-plane lattice depth and $E_R = \hbar^2 \pi^2 / (2m a_{\mathrm{latt}}^2)$ the recoil energy. The tighter $z$ confinement yields a 
larger gap $\lambda\,\hbar\omega_{\mathrm{ho}}$, justifying the quasi-2D reduction.

\begin{figure}
    \centering
    \includegraphics[width=1.0\linewidth]{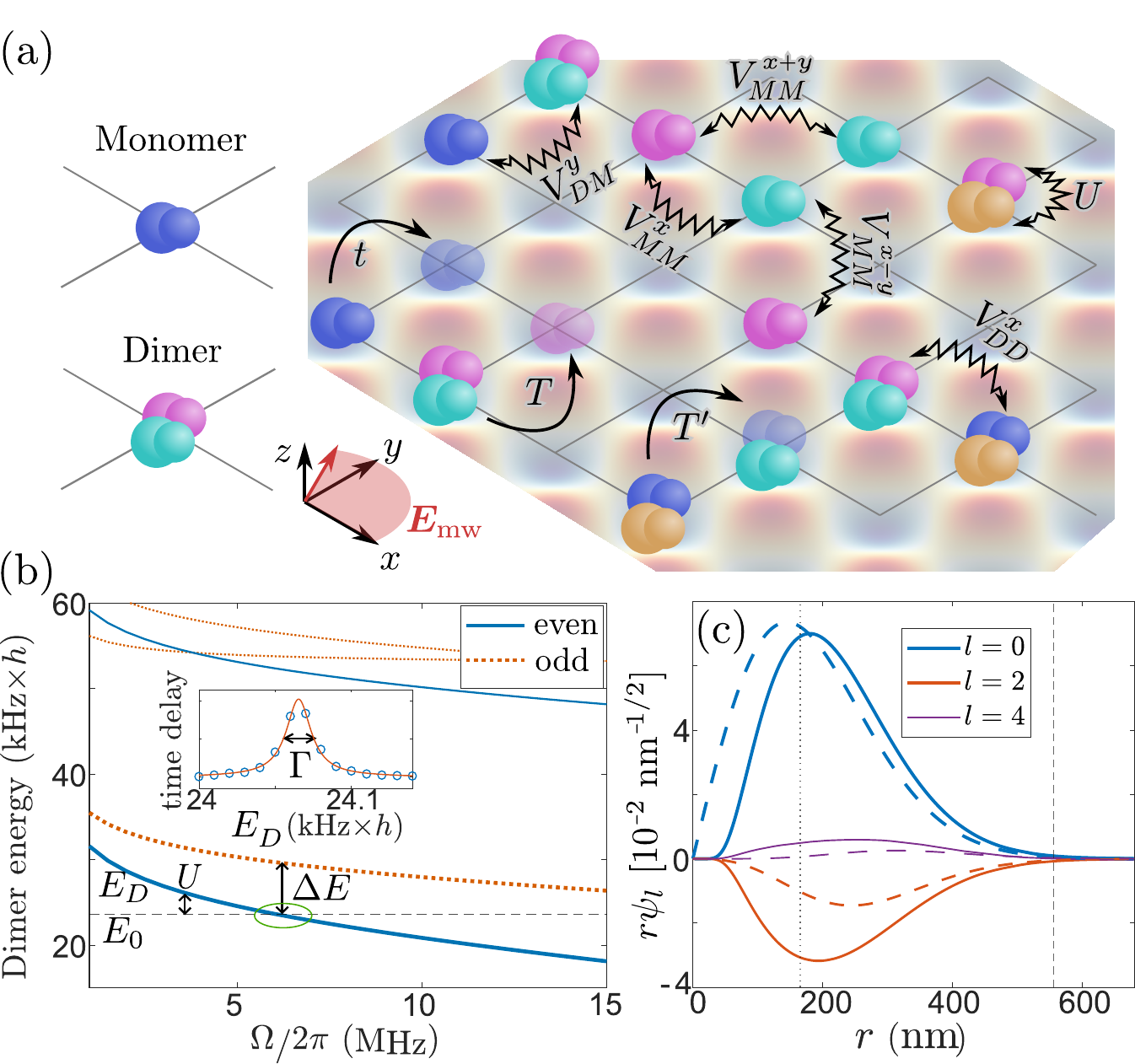}
    \caption{(a)~In a 2D lattice, dipolar molecules tunnel, form on-site dimers, and interact off site. (b)~Energy spectrum of two dipolar molecules in a single lattice site for even (solid lines) and odd (dashed lines) parities. The inset shows the typical Lorentzian time-delay profile of the finite-lifetime dimer. The solid line fits the numerical result (circles). (c)~Comparison of the on-site dimer-state wavefunction in terms of partial-wave angular momentum $l$, with (solid lines) and without (dashed lines) interaction. The dotted and dashed vertical lines indicate the $a_{\rm ho}$ and $a_{\rm latt}$, respectively. (b), (c) employ $a_{\rm latt}=550$ nm, $V_{0,xy}/E_R=5.1$, and $V_{0,z}=4V_{0,xy}$, (c) uses $\Omega = 2 \pi \times 6$ MHz.}
    \label{fig:sch}
\end{figure}

The conventional single-band Hubbard reduction assigns an on-site interaction energy $U_{\mathrm{bare}} = \langle w_i w_i | V_{\text{sh}}(\bm{r}_1-\bm{r}_2) | w_i w_i \rangle$ to each doubly occupied site. However, the strong repulsive core of the microwave-shielded potential pushes this energy to the MHz scale, far above $\hbar\omega_{\mathrm{ho}} \sim h \times 10$\, kHz. Double occupancies therefore lie outside the low-energy manifold, and the bare Hubbard description breaks down. Although the bare double occupation is excluded, an $s$-wave field-linked lattice-dimer state formed by an inter-spin pair with energy $E_D$ and even parity \cite{fld_note} can enter into the lowest-band lattice model as it is tuned to cross the two-monomer asymptotic energy, $E_0 = 2\hbar\omega_{\mathrm{ho}}$ (given $\omega_{z}=2\omega_{x,y}=2\omega_{\rm ho}$), via increasing $\Omega$, see Fig.~\ref{fig:sch}(b). The lattice dimer acquires linewidth $\hbar \Gamma$, that we calculate by solving the harmonically confined two-body system (see inset of Fig.~\ref{fig:sch}(b)), through inelastic decay into lower-lying microwave-dressed scattering channels~\cite{Chen:2024TM}, see the Supplemental Material~\cite{sm}. In this regime, a monomer can resonantly tunnel onto an adjacent occupied site to form a dimer, at the cost of the on-site energy $U = E_D - E_0$, depending on the dimer energy, $E_{\rm D}$.
The minimal energetic separation from other dimer and higher-body (not shown) states is $\Delta E \sim h \times 10$\, kHz (for the lowest odd-parity state), see Fig.~\ref{fig:sch}(b); thus they are energetically prohibited. 

The low-energy Hilbert space contains configurations where each site is empty or hosts either a monomer $\hat{c}^\dagger_{j;\alpha}$ (with spin index $\alpha$), or a dimer $\hat{d}^\dagger_{j;\alpha,\beta}$ (with $\alpha \neq \beta$, as required by fermionic statistics and even parity of the lattice-dimer state). The dimer wavefunction contains several even partial waves, dominated by the $s$ wave component, as shown in Fig.~\ref{fig:sch}(c).

Because the shielded interaction is nuclear-spin independent, the effective lattice model commutes with the SU($N$) generators~\cite{sm}. The most general SU($N$)-symmetric Hamiltonian neglecting interaction-induced tunneling effects in this projected basis reads
\begin{align}
\hat{H}_{\mathrm{eff}} = \hat{H}_t + \hat{H}_U + \hat{H}_T + \hat{H}_V,
\end{align}
where
$
\hat{H}_t = -t \sum_{\langle i,j\rangle,\alpha} \left( \hat{c}^\dagger_{i;\alpha}\,\prod_{\beta}(1 - \hat{n}_{i;\beta})\,\hat{c}_{j;\alpha} + \mathrm{h.c.} \right)
$
describes projected monomer tunneling with amplitude $t$. The Spa{\l}ek projector $\prod_{\beta}(1 - \hat{n}_{i;\beta})$ \cite{Chao1978canonical} is added to remove bare doublons.
The second term,
$
\hat{H}_U = U \sum_i \sum_{\alpha \neq \beta} \hat{d}^\dagger_{i;\alpha,\beta}\,\hat{d}_{i;\alpha,\beta},
$
assigns the on-site energy $U$ to each on-site dimer \cite{sm}, thereby playing the role of the Hubbard interaction. 
The third term,
\begin{align}
\hat{H}_T = &-T \sum_{\langle i,j\rangle} \sum_{\alpha \neq \beta} \left( \hat{d}^\dagger_{i;\alpha,\beta}\,\hat{c}_{i;\alpha}\,\hat{c}_{j;\beta} + \mathrm{h.c.} \right) \notag \\
&-T'\sum_{\langle i,j\rangle} \sum_{\beta \neq \alpha,\gamma \neq \alpha} \left(\hat{d}_{i; \alpha, \beta}^{\dagger} \hat{c}_{j; \gamma}^{\dagger} \hat{c}_{i; \beta} \hat{d}_{j; \alpha, \gamma}+\mathrm{h.c.} \right),
\end{align}
contains two tunneling terms. The $T$ term converts two monomers on different sites into an on-site dimer and vice versa. The $T'$ term transfers one molecule from a dimer to a neighboring monomer site, forming a new dimer. Here, we consider only the nearest-neighbor (${\langle i,j\rangle}$) processes and neglect the dimer tunneling term (see Refs.~\cite{Jurgensen2014,Dutta2015,Chanda:2025} for discussions of dimer tunneling). 
Finally,
\begin{align}
\hat{H}_V &= \sum_{\sigma=x,y,x\pm y}\left( \frac{V_{MM}^{\sigma}}{2}\sum_{\langle i,j\rangle_{\sigma}} \sum_{\alpha,\beta} \hat{c}_{j;\beta}^{\dagger} \hat{c}_{i;\alpha}^{\dagger}\hat{c}_{i;\alpha}\hat{c}_{j;\beta} \notag \right. \\
&+ V_{DM}^{\sigma}\sum_{\langle i,j\rangle_{\sigma}} \sum_{\alpha<\beta, \gamma} \hat{c}_{j;\gamma}^{\dagger} \hat{d}_{i;\alpha, \beta}^{\dagger} \hat{d}_{i;\alpha, \beta}\hat{c}_{j;\gamma}   \notag \\
&+\left. \frac{V_{DD}^{\sigma}}{2} \sum_{\langle i,j\rangle_{\sigma}} \sum_{\alpha<\beta, \gamma < \delta} \hat{d}_{j;\gamma, \delta}^{\dagger} \hat{d}_{i;\alpha, \beta}^{\dagger} \hat{d}_{i;\alpha, \beta}\hat{d}_{j;\gamma, \delta} \right)
\end{align}
collects off-site monomer-monomer, monomer-dimer, and dimer-dimer density couplings from the long-range dressed dipolar potential. Their strength and anisotropy depend on lattice geometry and dipole orientation relative to the lattice. In the above expression, we restrict the off-site interaction to be spin-independent and between nearest-neighbor (${\langle i,j\rangle}_{x,y}$) or diagonal nearest-neighbor (${\langle i,j\rangle}_{x \pm y}$) pairs. Spin-exchange and other off-site interactions are negligible, see \cite{sm}.

The extended-Hubbard parameters are extracted from microscopic monomer and dimer physics. The dimer--monomer-pair and dimer--monomer exchange couplings are given by the matrix elements
of the \textit{ab initio} single-molecule Hamiltonian.
The remaining $\hat{H}_V$ terms arise from the long-range dipolar interaction tail overlaps. Together, $t$, $U$, $T$, $T'$, $V^{\sigma}_{MM}$, $V^{\sigma}_{DM}$ and $V^{\sigma}_{DD}$ fully parametrize the extended SU($N$)-symmetric lattice Hamiltonian~\cite{sm}. In the following, we discuss the two experimental knobs controlling these parameters: the lattice depth $s=V_0/E_R$, which sets the single-particle band structure and tunneling amplitude $t$, and the microwave Rabi frequency $\Omega$, which determines the shielded two-body interaction and the resulting dimer properties.
\begin{figure}
    \centering
    \includegraphics[width=1.0\linewidth]{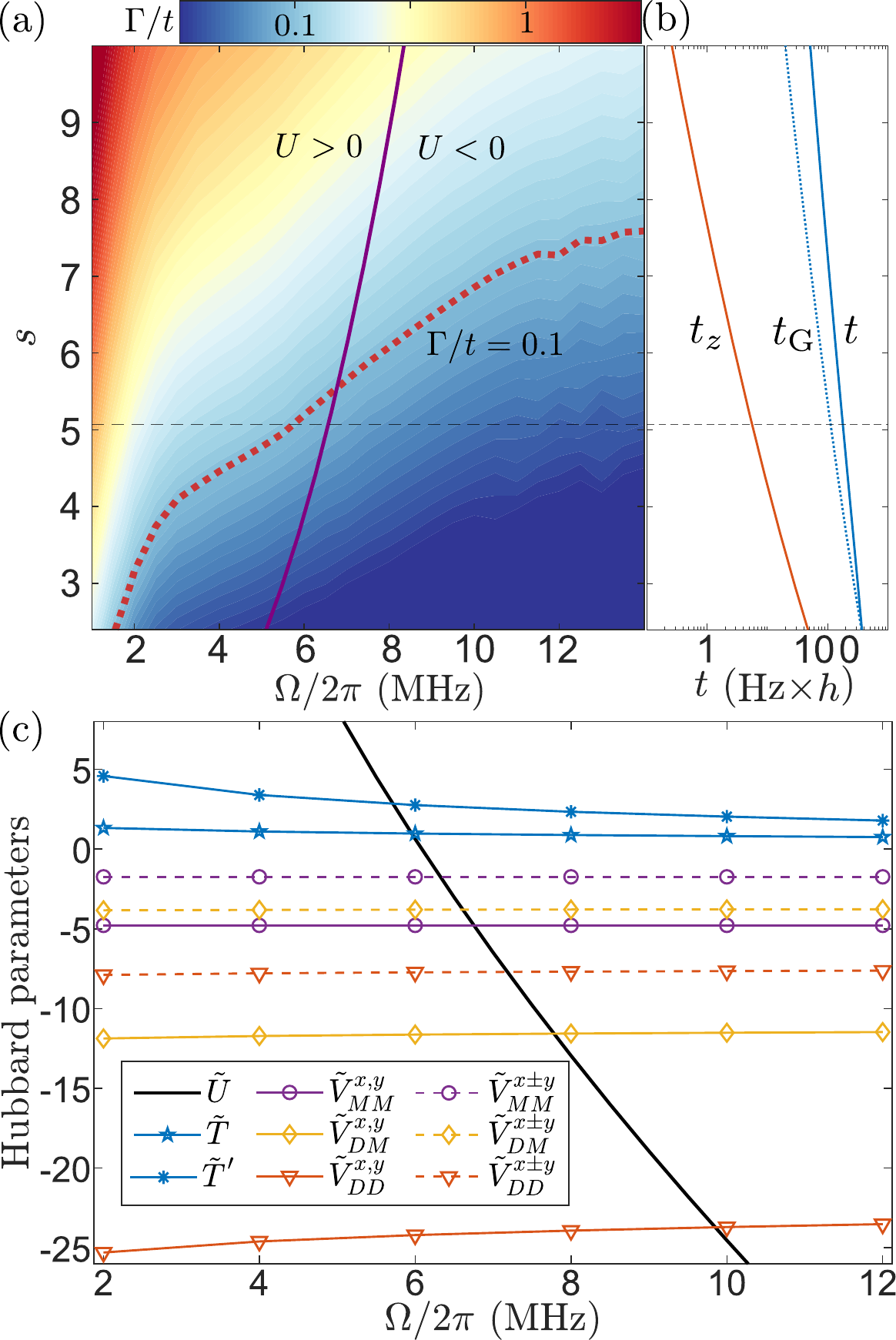}
    \caption{(a) The ratio of the linewidth of the dimer $\Gamma$ to the hopping parameter $t$; the lattice parameters are $a_{\rm latt}=550$~nm with $V_{0,z}=4V_{0,xy}$. (b) Hopping parameter $t$ in $x,y$ direction and $t_z$ in $z$ direction versus $s$. The dotted line displays the hopping parameter $t_{\rm G}$ in $x,y$ direction within the Gaussian approximation. (c) Dimensionless Hubbard parameters, defined as the ratio of the corresponding Hubbard parameter to $t_{\rm G}$, versus the microwave Rabi frequency. The horizontal dashed lines in (a) and (b) indicates $s=5.1$, at which the results of (c) are calculated. 
    }
    \label{fig:gam}
\end{figure}

To identify the experimentally accessible operating regime of the proposed Hubbard model, we analyze the relevant parameter space (Fig.~\ref{fig:gam}(a)). We adopt a conservative threshold of $\Gamma/t = 0.1$ (dotted red contour), below which the field-linked dimer is sufficiently long-lived to participate coherently in the lattice dynamics. Strong monomer-dimer hybridization is expected in the vicinity of the resonance condition $U(s,\Omega) = 0$ (solid purple curve), where the field-linked dimer crosses the lowest two-monomer threshold.
Crucially, a broad region of parameter space satisfies both $\Gamma/t < 0.1$ and proximity to the $U = 0$ line, showing that resonant monomer-dimer physics can be accessed under
experimentally realizable conditions. Figure~\ref{fig:gam}(b) compares the tunneling $t$ in $x,y$-direction with $t_z$ in $z$-direction at $\lambda=2$. The condition of $t_z \ll t$ reflects the quasi-2D nature of the dynamics. Here we provide two values of the tunneling parameter: $t$ extracted from the exact Wannier calculation \cite{Marzari1997Maximally} and $t_{\rm G}$ stemming from a Gaussian approximation as a reference for the following discussion \cite{sm}.

Along a representative cut through the accessible parameter space (horizontal dashed line in Fig.~\ref{fig:gam}(a)), we examine how the Hubbard parameters evolve with the microwave coupling $\Omega$ in Fig.~\ref{fig:gam}(c).
The tunneling amplitude $t$, set by the lattice depth alone, is constant along this trajectory, whereas the other parameters are presented as dimensionless ratios with respect to it. Since the Hubbard parameters are extracted from monomer and dimer states within the Gaussian approximation, we self-consistently normalize them with respect to $t_G$ instead of $t$, see also \cite{sm}. This is indicated by adding a tilde to the corresponding symbol, e.g. $\tilde{U} = U/t_{\rm G}$.
We found that the on-site dimer energy $U$ is strongly sensitive to the microwave dressing, where it decreases with increasing $\Omega$ and crosses zero. The monomer-dimer $T$ and inter-site dimer $T'$ conversion amplitudes vary weakly with $\Omega$, reflecting the robust spatial overlap between the dimer wavefunction and the monomer orbitals on neighboring sites.

The long-range character of the dressed dipolar interaction generates sizable extended couplings. The nearest neighbor monomer-monomer ($V_{MM}^{x,y}$), monomer-dimer ($V_{DM}^{x,y}$), and dimer-dimer ($V_{DD}^{x,y}$) density interactions, together with their diagonal counterparts, shown in Fig.~\ref{fig:gam}(c), are negative and typically larger than $T$ across all $\Omega$ explored. The diagonal nearest-neighbor interactions are generally weaker than the nearest-neighbor interactions owing to the larger particle separation. These off-site interactions originate from the long-range dipolar interaction tail of $V_{\text{sh}}$ and are largely insensitive to the short-range dimer structure \cite{sm}. 
The off-site interaction terms analyzed above compete with the tunneling and dimer--monomer(-pair) coupling processes and thus can impose charge density orders of different flavor \cite{Su2023dipolar}.
Nevertheless, we will show below that a high degree of tunability allows the interaction coefficients to become comparable to $t$, $T$ and $T'$, and even to cross zero.

Taken together, Fig.~\ref{fig:gam} demonstrates that a broad and experimentally accessible region of parameter space simultaneously satisfies three key conditions: (a) the dimer lifetime exceeds the lattice tunneling timescale ($\Gamma/t \ll 1$), (b) the detuning $U$ can be tuned continuously through zero, and (c) $U$ can be tuned to compete and interplay with the tunneling dynamics characterized by $t$, $T$, and $T'$, as well as, the off-site interactions. Within this regime, the system realizes a stable, strongly interacting SU($N$)-symmetric lattice gas in which monomer tunneling, coherent dimer formation, and long-range dipolar couplings are simultaneously relevant.


\begin{figure*}
    \centering
    \includegraphics[width=1.0\linewidth]{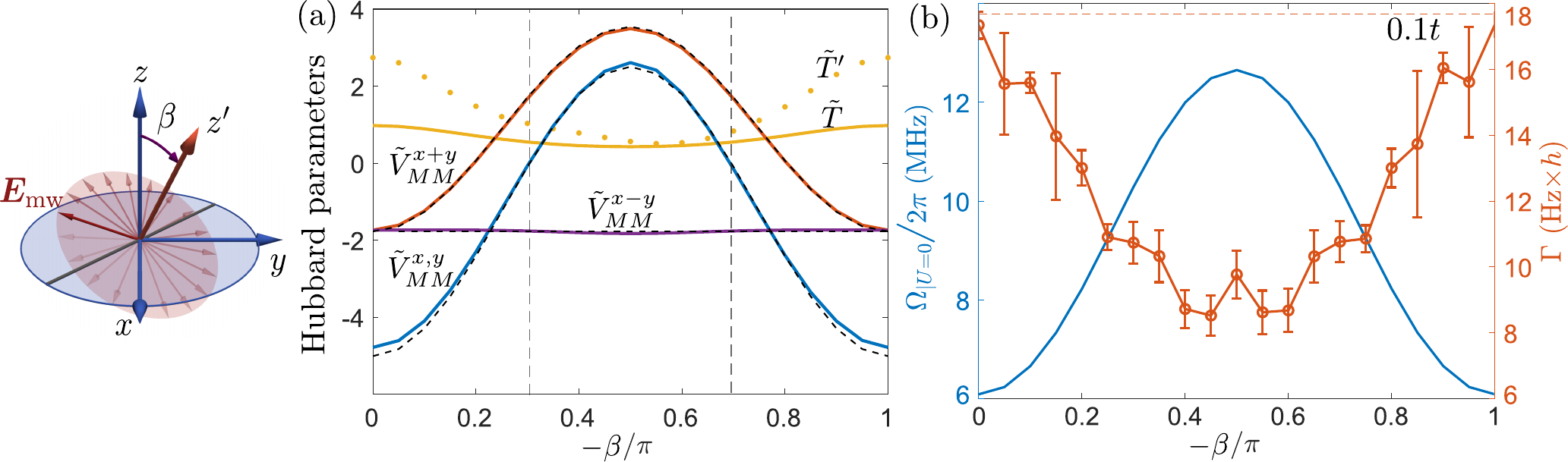}
    \caption{ (a) The dimensionless  hopping parameter and monomer-monomer nearest-neighbor and diagonal-nearest-neighbor interactions obtained by rotating the microwave field with the Euler angles of ($\alpha=\pi/4$, $\beta$, $\gamma=-\pi/4$) relative to the lattice frame.
    (b) The Rabi frequency at which the on-site energy crosses the energy of the non-interacting monomer pair and corresponding dimer linewidth versus the rotation angle $\beta$. The horizontal dashed line indicates the value of $0.1 t$ on the right $y$-scale. The lattice parameters are chosen as $a_{\rm latt}=550$ nm with $V_{0,xy}/E_R=5.1$ and $V_{0,z}=4V_{0,xy}$. The dashed lines close to the numerical results (solid lines) of $\tilde{V}_{MM}^{x,y}$,$\tilde{V}_{MM}^{x+y}$ and $\tilde{V}_{MM}^{x-y}$ display the prediction from the point-particle approximation \cite{Dutta2015,Zhang2022}. }
    \label{fig:angle}
\end{figure*}

An additional experimental knob arises from the orientation of the microwave polarization axis relative to the lattice. Because the dressed dipole-dipole interaction is inherently anisotropic, rotating the field with respect to the lattice frame modifies the off-site couplings and the on-site dimer properties while preserving the underlying SU($N$) symmetry. We parametrize this rotation by the Euler angles $(\alpha,\beta,\gamma)$ that connect the two frames. We fix $\alpha = \pi/4$ and $\gamma = -\pi/4$, while scanning the tilt angle $\beta$, see Fig. \ref{fig:angle}.

The nearest-neighbor $V^{x,y}_{MM}$ and diagonal nearest-neighbor $V_{MM}^{x+y}$ interactions both vary strongly with the orientation angle $\beta$, sweeping through zero, as shown in Fig.~\ref{fig:angle}(a).
Notably, $V_{MM}^{x,y}$ crosses zero near $\beta/\pi \approx 0.3$, while the diagonal coupling $V_{MM}^{x+y}$ remains sizable and positive at the same angle. In contrast, the anti-diagonal nearest-neighbor $V_{MM}^{x-y}$ interaction depends only weakly on $\beta$. Nevertheless, the behaviors of $V^{x,y}_{MM}$ and $V_{MM}^{x\pm y}$ is well captured by a sinusoidal curve within the point-particle approximation \cite{Dutta2015,Zhang2022}. We find that the dimer-monomer and dimer-dimer interactions share
a similar angular dependence to the monomer-monomer interaction~\cite{sm}.
This demonstrates that the nearest-neighbor interaction can be selectively switched off by geometric tuning alone, while tunneling, on-site fluctuations and other long-range couplings remain active. Such selective suppression opens access to regimes of the extended Hubbard model in which diagonal or further-neighbor interactions dominate over nearest-neighbor ones, a situation difficult to achieve in other settings.

Figure~\ref{fig:angle}(b) confirms that this tunability is fully compatible with dimer stability. It displays the Rabi frequency $\Omega|_{U=0}$ at which the dimer resonance condition $U = 0$ is met, together with the corresponding decay linewidth $\Gamma$, both as functions of $\beta$. The resonance frequency varies smoothly between roughly $6$ and $13$~MHz, confirming that the dimer can be brought into resonance independently of orientation. The decay linewidth remains $\sim 8$-$18\;\mathrm{Hz}\times h$ throughout, well below the tunneling parameter $t = 177\;\mathrm{Hz}\times h$ at the chosen lattice parameters ($a_{\mathrm{latt}} = 550\;\mathrm{nm}$, $V_{0,xy}/E_R = 5.1$, $V_{0,z} = 4V_{0,xy}$), ensuring $\Gamma/t \ll 1$ for all angles. The system therefore remains deep in the coherent regime across the full orientational scan, including the angles where $V_{MM}^{x,y}$ vanishes.

Geometric rotation thus complements the Rabi frequency as a practical tuning knob, providing independent control over the magnitude, sign, and spatial structure of the off-site interactions. In particular, the ability to continuously suppress the nearest-neighbor coupling while maintaining stable dimer formation and finite diagonal interactions makes it possible to navigate between qualitatively distinct regimes of the SU($N$)-symmetric lattice model within a single experimental platform.


We have derived an SU($N$)-symmetric extended Hubbard model for
microwave-shielded fermionic molecules in an optical lattice, where on-site interaction can be
synthesized by coupling to a long-lived field-linked lattice dimer state. We demonstrate the experimental tunability of this Hamiltonian 
in terms of the on-site and off-site interactions, tunable via the microwave Rabi frequency and field orientation, respectively. 
Our approach can be generalized to bosonic \sun-symmetric Hubbard models, for instance, for ${}^{23}$Na${}^{39}$K \cite{Vorges2020ultracold,Li2026}, opening up opportunities to explore a broader class of Hubbard physics than are accessible with atomic platforms.


Possible applications include dimer superfluidity, resonant pairing mechanisms, and lattice models with tunable anisotropic interactions. The
dimer lifetime is already compatible with
coherent lattice dynamics, but improvements might be possible using double
microwave shielding~\cite{Karman2025double,Deng2025} or a combination of
circular microwave dressing and static electric
fields~\cite{Gorshkov2008Suppression,Ho2026tuning,Wang2026Bound}.
A parallel study shows, for double-shielded molecules, flexible tuning of on-site and off-site interactions for various dipolar molecules and how these interactions can serve as a tool to characterize effective hole-hole and other inter-dopant interactions~\cite{Perez2026Hubbard}. The present work demonstrates how coupling to a lattice dimer enables an \sun-symmetric extended Hubbard model with independently tunable on-site and off-site interactions.

\begin{acknowledgements}
J.-L. L. and G. M. K. were funded by the Austrian Science Fund (FWF) [10.55776/F1004]. 
R. A. received funding from the Austrian Academy of Science ÖAW grant No. PR1029OEAW03. 
A. S. acknowledges support through the ERC grant UltraMeDiQs (project No.\ 101219560). Funded by the European Union. Views and opinions expressed are however those of the authors only and do not necessarily reflect those of the European Union or the European Research Council Executive Agency. Neither the European Union nor the granting authority can be held responsible for them.
K.R.A.H. is supported in part by the W. M. Keck Foundation (Grant No. 995764) and K.R.A.H. is supported in part by National Science Foundation (NSF DGE-2346014). A.S. and K.R.A.H. performed part of this work at the Aspen Center for Physics, which is supported in part by the National Science Foundation (PHY-1066293).
\end{acknowledgements}
\clearpage
\newpage
\section{Supplemental Material}
\subsection{Two-body problem in a lattice site}

Under the harmonic approximation, the two-molecule problem within a single lattice site is described by
\begin{equation}
\hat H_2
=
-\frac{\hbar^2}{2m}\nabla_1^2
-\frac{\hbar^2}{2m}\nabla_2^2
+\frac{1}{2}m\omega^2(\bm{r}_1^2+\bm{r}_2^2)
+
V_{\text{sh}}(\bm{r}_2-\bm{r}_1),
\label{eq:H2}
\end{equation}
where $V_{\text{sh}}(\mathbf r)$ is the microwave-dressed interaction potential
between two molecules in the repulsive shielding channel. We consider in this work an anisotropic trap with $\omega=(\omega_{\rm ho},\omega_{\rm ho}, 2\omega_{\rm ho})$ in ($x,y,z$) direction.

\begin{figure}
    \centering
    \includegraphics[width=1.0\linewidth]{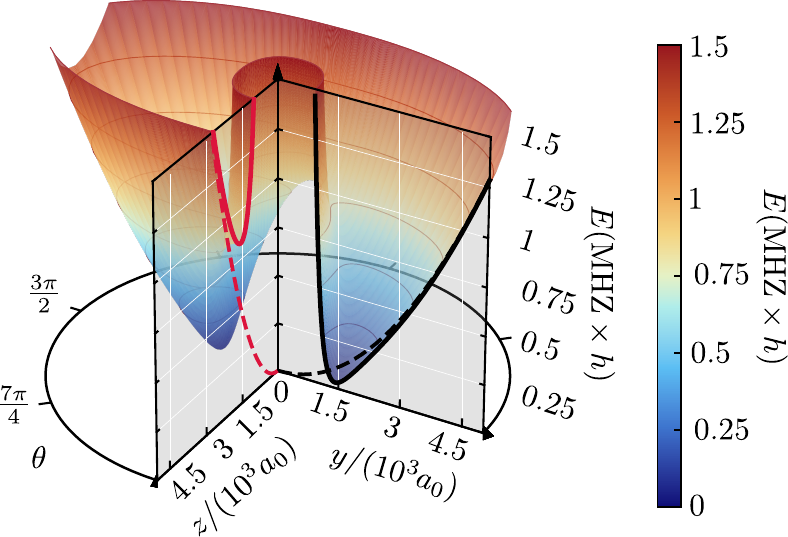}
    \caption{Total potential energy in the relative frame $\frac14 m \omega^2 {\bm r}^2 + V_{\rm sh}({\bm r})$. The dashed lines in the $zE$ and $yE$ planes indicate the harmonic-oscillator potential in comparison to the full potential (solid line) in the corresponding planes. We consider ${}^{23}$Na${}^{40}$K with $V_{0,xy}/E_{\rm R} = 5.1$, $V_{0,z}=4V_{0,xy}$, $a_{\rm latt} = 550$ nm, $\Omega = 2 \pi \times 6$ MHz and $\delta = 0$.}
    \label{fig:potential}
\end{figure}

Introducing center-of-mass and relative coordinates,
$
\bm{R}=\frac{\bm{r}_2+\bm{r}_1}{2},
\,
\bm{r}=\bm{r}_2-\bm{r}_1 ,
$
the Hamiltonian separates as
$
\hat H_2
=
\hat H_{\mathrm{cm}}
+
\hat H_{\mathrm{rel}},
$
with
\begin{align}
\hat H_{\mathrm{cm}}
&=
-\frac{\hbar^2}{4m}\nabla_R^2
+
\frac{1}{2}(2m)\omega^2\bm{R}^2 ,
\\
\hat H_{\mathrm{rel}}
&=
-\frac{\hbar^2}{m}\nabla_r^2
+
\frac{1}{4}m\omega^2\bm{r}^2
+
V_{\text{sh}}(\bm{r}).
\end{align}
The spectrum of the relative Hamiltonian $\hat H_{\mathrm{rel}}$ contains both scattering states and bound or quasi-bound states. The field-linked dimer states of interest correspond to bound solutions of the relative
Hamiltonian
$
\hat H_{\mathrm{rel}}\psi_D(\mathbf r)
=
E_D\psi_D(\bm{r}),
$
whose wavefunctions are localized at intermolecular separations set by the
characteristic length scale of the dressed dipolar potential. The total potential in relative coordinates is shown in Fig.~\ref{fig:potential}. To get the wavefunction, we solve the equation of relative motion using the DVR method \cite{Willner2004}. We note that a more accurate description of the on-site two-body problem requires going beyond the harmonic approximation, thereby coupling the center-of-mass motion to the relative degree of freedom, as is implemented in Ref. \cite{karman2026}. 

As the microwave Rabi frequency $\Omega$ is varied, the energy of a
field-linked state can be tuned relative to the trap spectrum.
Figure~\ref{fig:sch}(b) illustrates this mechanism where we show that the dimer energy
$E_D(\Omega)$ approaches the energy of the lowest trap state
$
E_0=2\hbar\omega_{\mathrm{ho}},
$
and eventually crosses it. Near this crossing the two-body spectrum exhibits a
well-isolated resonance resulting from hybridization between the scattering
channel of two shielded molecules and the field-linked dimer state
(see Fig.~\ref{fig:sch}), and we can define the on-site detuning parameter as $U=E_D-E_0.$

The lifetime of the resonance can be characterized by the Wigner time delay associated with the scattering matrix $S$, which encodes both the elastic phase shift (setting the resonance position) and the decay probability (determining the width). We extract the $S$-matrix from the solution of the two-body Schr\"{o}dinger equation of $\hat H_{\mathrm{rel}}$, with additional low-lying microwave-dressed potentials included and absorbing boundary conditions imposed \cite{Li2026}.
Due to the non-unitarity of the $S$ in the presence of the absorbing boundary condition, the time delay is calculated from a more general formula \cite{Chen2021}:
\begin{equation}
\tau (E) =-i\hbar{\rm Tr}\left(S^{\dagger}\frac{d S}{dE}\right).
\end{equation}
Near a resonance, the time delay follows a Lorentzian profile
\begin{equation}
\tau(E)
=
\frac{\hbar\Gamma}{(E-E_D)^2+(\Gamma/2)^2}.
\end{equation}
The width $\Gamma$ determines the decay rate of the metastable dimer state
and thus its lifetime
$
\hbar/\Gamma.
$ 

\subsection{SU(N) matrix elements}
\subsubsection{Hamiltonian definitions}
The effective Hubbard Hamiltonian is derived by projecting the {\it ab initio} Hamiltonian to the low energy sector corresponding to monomers and dimers. To evaluate, the possible terms that can appear in this projection we have to consider that the {\it ab initio} Hamiltonian, $\hat{H}$, is composed of a single particle Hamiltonian, $\hat{H}_{1}$, corresponding to the kinetic energy of the molecules and their potential and a two-body Hamiltonian, $\hat{H}_2$, corresponding to the interaction among the particles. With respect to the spin degree-of-freedom we assume that $\hat{H}_1$ and $\hat{H}_2$ are spin-independent.

\subsubsection{Number-state basis}
The first step in the projection is to consider the most generic state on the lattice composed solely of monomers and dimers. This can be written in terms of the number states characterized by a given occupation of each site. For $N$ spin states there are $1 + \frac{N(N+1)}{2}$ possible states on each lattice site, the first and reference state is the vacuum, $| n_j = 0 \rangle \equiv | 0 \rangle$, where the site is not occupied. Beyond this, there are $N$ further states corresponding to monomer occupation $|n_{j} = \alpha \rangle \equiv \hat{c}_{j;\alpha}^{\dagger} | 0 \rangle$ and $\frac{N(N-1)}{2}$ states corresponding to dimer occupation $| n_{j} = \{ \alpha, \beta \} \rangle \equiv \hat{d}_{j; \alpha, \beta}^{\dagger} | 0 \rangle$. In order to avoid double counting we have enforced $\alpha < \beta$ for the spin indices of the dimer states. Then any number state can be written as $| \vec{n} \rangle = |n_1, n_2, \dots, n_M \rangle$, where $M$ is the number of sites and $n_j \in \{ 0, 1, \dots, N \} \cup \{\{\alpha, \beta\}: \alpha, \beta \in \{1, \dots, N\} \text{ and } \alpha < \beta \}$. 

\subsubsection{Single-body Hamiltonian, $\hat{H}_1$, matrix elements}
Let us now consider the matrix elements of the Hamiltonian $\hat{H}$ in this number state basis, provided the properties of $\hat{H}_{1}$ and $\hat{H}_2$. Since $\hat{H}_1$ is a single particle operator, it is required that it only allows for non-zero matrix elements between states differing by the position of at most one particle. Given also that this term is spin-independent, the moved particle should have the same spin as before it moves. This is possible in four cases: (a) one monomer moved to an unoccupied site, (b) a monomer moved to a site occupied by another monomer with different spin and formed a dimer or (c) vice versa, finally, (d) a particle from a dimer state moved to another site occupied by a monomer, thus creating a new dimer and leaving behind a monomer in its original site. These imply that the action of $\hat{H}_1$ on a number state can be written as
\begin{equation}
\begin{split}
  \hat{H}_1 | \vec{n} \rangle &= \left[ \sum_{i} \epsilon^{(1)}_{i}(n_{i}) \right] |\vec{n}\rangle + \sum_{i \neq j} \sum_{\alpha} t_{i, j} \hat{c}_{i; \alpha}^{\dagger} \hat{c}_{j; \alpha} | \vec{n} \rangle \\
  &+\sum_{i \neq j} \sum_{\alpha} \sum_{\beta \neq \alpha} \left( T_{i, j} \hat{d}_{i; \alpha, \beta}^{\dagger} \hat{c}_{i; \beta} \hat{c}_{j; \alpha} + {\rm h.c.} \right) | \vec{n} \rangle \\
  &+\sum_{i \neq j} \sum_{\alpha} \sum_{\beta \neq \alpha} \sum_{\gamma \neq \alpha} T_{i, j}' \hat{d}_{i; \alpha, \beta}^{\dagger} \hat{c}_{j; \gamma}^{\dagger} \hat{d}_{j; \alpha, \gamma}\hat{c}_{i; \beta} | \vec{n} \rangle.
\end{split}
\label{single-particle-hamiltonian}
\end{equation}
Notice that due to the single particle nature of the Hamiltonian term, its diagonal matrix elements are only allowed to depend in a linear fashion to the occupation of each site, since the state in different sites is related by a tensor product within the number state basis. Notice that while the additional constraint $\epsilon_{i}^{(1)}(\{\alpha, \beta\}) = \epsilon_{i}^{(1)}(\alpha) + \epsilon_{i}^{(1)}(\beta)$ is tempting, it is not enforced here since we consider that the dimer state is not necessarily a tensor product of monomer states. Since the Hamiltonian is spin-independent, it is allowed to act only on one of the spin-indices, here $\alpha$. Other involved spin-indices, appearing when dimers are involved, are constrained to be different than $\alpha$ in order for the dimer state to exist. In addition, recall that the indices appearing in dimer operators have to be ordered, we have not implemented this in Eq.~\eqref{single-particle-hamiltonian} to make it less cumbersome. However, in applications ordering should be enforced.

By using the \textit{ab-initio} Hamiltonian, the couplings $\epsilon^{(1)}_{i}(n_i)$, $t_{i, j}$, $T_{i,j}$ and $T_{i, j}'$ can be calculated in terms of the matrix elements
\begin{equation}
\begin{split}
\epsilon_i(n_i = 0) &= 0, \\
\epsilon_i(n_i = \alpha) &= \langle 0 | \hat{c}_{i; 1} \hat{H}_1 \hat{c}_{i; 1}^{\dagger}| 0 \rangle, \\
\epsilon_i(n_i = \{\alpha, \beta\}) &= \langle 0 | \hat{d}_{i; 1, 2} \hat{H}_1 \hat{d}_{i; 1, 2}^{\dagger}| 0 \rangle, \\
t_{i,j} &= \langle 0 | \hat{c}_{i; 1} \hat{H}_1 \hat{c}_{j; 1}^{\dagger}| 0 \rangle, \\
T_{i,j} &= \langle 0 | \hat{d}_{i; 1,2} \hat{H}_1 \hat{c}_{j; 1}^{\dagger} \hat{c}_{i; 2}^{\dagger}| 0 \rangle, \\
T'_{i,j} &= \langle 0 | \hat{d}_{i; 1,2} \hat{c}_{j, 2} \hat{H}_1 \hat{d}_{j; 1, 2}^{\dagger} \hat{c}_{i; 2}^{\dagger}| 0 \rangle,
\end{split}
\end{equation}
where we used the spin-independence of $\hat{H}_1$ to reduce the involved spin-states to the minimal set, $\alpha \in \{1,2\}$. In applications, the remaining spin indices can be used to verify the correctness of the calculation.

\subsubsection{Two-body Hamiltonian, $\hat{H}_2$: term definitions}
For the two-body interaction the considerations become more complicated. In this case in order for a matrix element to be non-zero there can be a difference in the particle occupation at most four different sites. This results to $31$ different classes of interaction terms summarized in the Tables \ref{tab:Dtab}, \ref{tab:MMtab}, \ref{tab:DMtab}, and \ref{tab:DDtab}.  Each class is given a label \(X:y \to Z\),
where \(X \in \{{ D}, { MM}, { DM}, { DD}\}\), with the symbol descriptions in Table \ref{tab:Xdescr},
\(y \in \{\varnothing, { d}, { d}^{2}, { mm}, { dm}, { dd} \}\), with the symbol descriptions in Table \ref{tab:ydescr},
\(Z \in \{\varnothing, { E}, { M}, { O}, { EE}, { E}^{2}, { EM}, { MM}\}\), with the symbol descriptions in Table \ref{tab:Zdescr}.
The class \(X:y \to Z\) corresponds to \({\rm desc}(X)\) interacting, resulting in \({\rm desc}(y)\) moving into \({\rm desc}(Z)\),
where \({\rm desc}(x)\) means substitute the symbol \(x\) by its description in the corresponding table.
The Tables \ref{tab:Dtab}, \ref{tab:MMtab}, \ref{tab:DMtab}, and \ref{tab:DDtab} are separated in terms of different \(X\).

\begin{table}[h]
\caption{\label{tab:Xdescr} Interacting particle class symbols, \(X\), and their descriptions, \({\rm desc}(X)\).}
\centering
\begin{tabular}{ll}
\(X\) & \({\rm desc}(X)\)\\
\hline
\({D}\) & the particles of a single dimer\\
\({MM}\) & one monomer and a different monomer\\
\({DM}\) & a particle of a dimer and a monomer\\
\({DD}\) & two particles of different dimers\\
\end{tabular}
\end{table}

\begin{table}[h]
\caption{\label{tab:ydescr} Particle moved due to interaction class symbols, \(y\), and their descriptions, \({\rm desc}(y)\).}
\centering
\begin{tabular}{ll}
\(y\) & \({\rm desc}(y)\)\\
\hline
\(\varnothing\) & no particle\\
\({d}\) & one particle of a dimer\\
\({d}^2\) & both particles of a dimer\\
\({mm}\) & two different monomers\\
\({dm}\) & a particle of a dimer and a monomer\\
\({dd}\) & two particles of different dimers\\
\end{tabular}
\end{table}

\begin{table}[h]
\caption{\label{tab:Zdescr} Sites particles moved into due to interaction class symbols, \(Z\), and their descriptions, \({\rm desc}(Z)\).}
\centering
\begin{tabular}{ll}
\(Z\) & \({\rm desc}(Z)\)\\
\hline
\(\varnothing\) & no other sites\\
\({E}\) & an empty site\\
\({M}\) & a site hosting a monomer\\
\({O}\) & the site of the particle that interacted \\
         & with the particle that moved\\
\({EE}\) & two different empty sites\\
\({E}^2\) & the same empty site\\
\({EM}\) & an empty site and a site hosting a monomer\\
\({MM}\) & two distinct sites hosting monomers\\
\end{tabular}
\end{table}

\subsubsection{Two-body Hamiltonian, $\hat{H}_2$: Operators and symmetry tensors}
Each class is associated with an operator $\hat{O}_{j_1, \dots, j_{N_s};\alpha_1, \dots, \alpha_{N_{\sigma}}}^{k_1, \dots, k_{N_s};\beta_1, \dots, \beta_{N_{\sigma}}}$, with $N_s \leq 4$ site and $N_{\sigma} \leq 6$ spin indices, composed of products of monomer and dimer creation and annihilation operators. To enforce the Hermiticity of the Hamiltonian, the two-body character and the SU($N$) invariance of the term this operator is contracted with a spin tensor $\hat{M}_{\alpha_1, \dots, \alpha_{N_{\sigma}}}^{\beta_1, \dots, \beta_{N_{\sigma}}}$. To preserve the SU($N$) symmetry the tensor $\hat{M}_{\alpha_1, \dots, \alpha_{N_{\sigma}}}^{\beta_1, \dots, \beta_{N_{\sigma}}}$ contains factors of
\begin{equation}
T^{\beta_ 1,\beta_ 2,\dots,\beta_{N_{\sigma}}}_{\alpha_ 1,\alpha_2,\dots,\alpha_{N_{\sigma}}}=\sum_{i=1}^{N_{\sigma}!}c_i\prod_{j=1}^{N_{\sigma}}\delta^{\beta_j}_{\alpha_{P_i(j)}},
\end{equation}
with $c_{i} \in \mathbb{C}$ expansion coefficients, $P_{i}(j)$ the $j$-th element of the $i$-th permutation of the set $\{1, 2, \dots, N_{\sigma}\}$ and $\delta_{\alpha}^{\beta}$ the Kronecker delta, ensuring that the spin of the particles does not change due to the interaction. Notice that the tensor $T^{\beta_ 1,\beta_ 2,\dots,\beta_{N_{\sigma}}}_{\alpha_ 1,\alpha_2,\dots,\alpha_{N_{\sigma}}}$ contains all coupling constants, $c_{i}$, which determine the strength of each interaction that are different for each class. In principle there are $N_{\sigma}!$ such parameters, but their number decreases by incorporating the additional symmetries characterizing the term. To enforce $s$-wave dimers composed of two fermions we need to antisymmetrize the indices of $T^{\beta_ 1,\beta_ 2,\dots,\beta_{N_{\sigma}}}_{\alpha_ 1,\alpha_2,\dots,\alpha_{N_{\sigma}}}$. This we indicate by placing the antisymmetrized indices in parenthesis. E.g. $T_{(\alpha, \beta)}^{\gamma, \delta} = \frac{1}{2}\left( T_{\alpha, \beta}^{\gamma, \delta} - T_{\beta, \alpha}^{\gamma, \delta}\right)$. To ensure that the interactions are two-body we further ensure that the spin of spectator particles, required for describing the dimer when a particle moves into an occupied site after interaction, does not change. For the case that a dimer transforms to monomers and vice-versa this is achieved via the tensor
\begin{equation}
\label{eq:2}
\Delta_{\alpha}^{\beta_1, \beta_2} = \delta_{\alpha}^{\beta_1}\left(1 - \delta_{\alpha}^{\beta_2} \right) + \delta_{\alpha}^{\beta_2}\left(1 - \delta_{\alpha}^{\beta_1} \right).
\end{equation}
For dimers that remain dimers after the interaction while one of their particles is considered spectator, e.g. within interactions the ${ DM}:{ m}\to{E}$ class, we define
\begin{equation}
\Delta_{\alpha_1, \alpha_2}^{\beta_1, \beta_2} = 1 -(1-\Delta_{\alpha_1}^{\beta_1, \beta_2}) (1-\Delta_{\alpha_2}^{\beta_1, \beta_2}).
\end{equation}
Finally, for self-adjoint classes we need to ensure that the total operator is equal to its conjugate and thus we need to symmetrize the corresponding $\hat{M}_{\alpha_1, \dots, \alpha_{N_{\sigma}}}^{\beta_1, \dots, \beta_{N_{\sigma}}}$, since this need to be done on a case by case basis we schematically indicate this process by the $C$ function in Tables~\ref{tab:Dtab}, \ref{tab:MMtab}, \ref{tab:DMtab}, and \ref{tab:DDtab}.

\begin{table*}[htbp]
\caption{\label{tab:Dtab} Dimer classes}
\centering
\begin{tabular}{llll}
class & operator & adjoint class & Spin term\\
\hline
\({ D}:\varnothing \to \varnothing\) & \(\hat{d}_{i; \gamma, \delta}^{\dagger}\hat{d}_{i; \alpha, \beta}\) & self & \(C\left(T_{(\alpha, \beta)}^{(\gamma,\delta)}\right)\)\\
\({ D}:{ d}\to{ E}\) & \(\hat{c}_{j;\gamma}^{\dagger} \hat{c}_{i;\delta}^{\dagger} \hat{d}_{i; \alpha, \beta}\) & \({ MM}:{ m} \to { O}\) & \(T_{(\alpha,\beta)}^{\gamma, \delta} \Delta_{\delta}^{\alpha, \beta}\)\\
\({ D}:{ d}\to{ M}\) & \(\hat{d}_{j;\delta, \epsilon}^{\dagger} \hat{c}_{i;\zeta}^{\dagger} \hat{d}_{i; \alpha, \beta} \hat{c}_{j;\gamma}\) & self & \(C\left(T_{(\alpha, \beta),\gamma}^{(\delta, \epsilon),\zeta} \Delta_{\gamma}^{\delta, \epsilon} \Delta_{\zeta}^{\alpha, \beta}\right)\)\\
\({ D}:{ d}^2\to{ E}^2\) & \(\hat{d}_{j;\gamma, \delta}^{\dagger} \hat{d}_{i; \alpha, \beta}\) & self & \(C\left(T_{(\alpha,\beta)}^{(\gamma, \delta)}\right)\)\\
\({ D}:{ d}^2\to{ EE}\) & \(\hat{c}_{k;\delta}^{\dagger} \hat{c}_{j;\gamma}^{\dagger} \hat{d}_{i; \alpha, \beta}\) & \({ MM}:{ mm} \to { E}^2\) & \(T_{(\alpha, \beta)}^{\gamma, \delta}\)\\
\({ D}:{ d}^2\to{ EM}\) & \(\hat{c}_{k;\zeta}^{\dagger} \hat{d}_{j; \delta, \epsilon}^{\dagger} \hat{d}_{i; \alpha, \beta} \hat{c}_{j, \gamma}\) & \({ DM}:{ dm}\to E^2\) & \(T_{(\alpha, \beta), \gamma}^{(\delta, \epsilon), \zeta} \Delta_{\gamma}^{\delta,\epsilon}\)\\
\({ D}:{ d}^2\to{ MM}\) & \(\hat{d}_{k;\eta, \theta}^{\dagger} \hat{d}_{j;\epsilon, \zeta}^{\dagger} \hat{d}_{i; \alpha, \beta} \hat{c}_{j; \gamma} \hat{c}_{k; \delta}\) & \({ DD}:{ dd} \to { E}^2\) & \(T_{(\alpha, \beta), \gamma, \delta}^{(\epsilon,\zeta), (\eta,\theta)} \Delta_{\gamma}^{\epsilon, \zeta} \Delta_{\delta}^{\eta, \theta}\)\\
\end{tabular}
\end{table*}

\begin{table*}[htbp]
\caption{\label{tab:MMtab} Monomer-monomer classes}
\centering
\begin{tabular}{llll}
class & operator & adjoint class & Spin term\\
\hline
\({ MM}:\varnothing \to \varnothing\) & \(\hat{c}_{j; \gamma}^{\dagger} \hat{c}_{i; \delta}^{\dagger} \hat{c}_{i; \alpha} \hat{c}_{j; \beta}\) & self & \(C\left(T_{\alpha, \beta}^{\gamma, \delta}\right)\)\\
\({ MM}:{ m}\to{ E}\) & \(\hat{c}_{k; \gamma}^{\dagger} \hat{c}_{i; \delta}^{\dagger} \hat{c}_{i; \alpha} \hat{c}_{j; \beta}\) & self & \(C\left(T_{\alpha, \beta}^{\gamma, \delta}\right)\)\\
\({ MM}:{ m}\to{ M}\) & \(\hat{d}_{k; \delta, \epsilon}^{\dagger} \hat{c}_{i; \zeta}^{\dagger} \hat{c}_{i; \alpha} \hat{c}_{j; \beta} \hat{c}_{k; \gamma}\) & \({ DM}:{ d}\to{ E}\) & \(T_{\alpha, \beta, \gamma}^{(\delta, \epsilon), \zeta} \Delta_{\gamma}^{\delta,\epsilon}\)\\
\({ MM}:{ m}\to{ O}\) & \(\hat{d}_{i; \gamma, \delta}^{\dagger} \hat{c}_{i; \alpha} \hat{c}_{j; \beta}\) & \({ D}:{ d}\to{ E}\) & \(T_{\alpha, \beta}^{(\gamma,\delta)} \Delta_{\alpha}^{\gamma, \delta}\)\\
\({ MM}:{ mm}\to{ EE}\) & \(\hat{c}_{l; \delta}^{\dagger} \hat{c}_{k; \gamma}^{\dagger} \hat{c}_{i; \alpha} \hat{c}_{j; \beta}\) & self & \(C\left(T_{\alpha, \beta}^{\gamma, \delta}\right)\)\\
\({ MM}:{ mm}\to{ E}^2\) & \(\hat{d}_{k; \gamma, \delta}^{\dagger} \hat{c}_{i; \alpha} \hat{c}_{j; \beta}\) & \({ D}:{ d}^2\to{ EE}\) & \(T_{\alpha, \beta}^{(\gamma, \delta)}\)\\
\({ MM}:{ mm}\to{ EM}\) & \(\hat{d}_{l; \epsilon, \zeta}^{\dagger} \hat{c}_{k; \delta}^{\dagger} \hat{c}_{i; \alpha} \hat{c}_{j; \beta} \hat{c}_{l; \gamma}\) & \({ DM}:{ dm}\to{ EE}\) & \(T_{\alpha, \beta, \gamma}^{\delta, (\epsilon, \zeta)} \Delta_{\gamma}^{\epsilon, \zeta}\)\\
\({ MM}:{ mm}\to{ MM}\) & \(\hat{d}_{l; \eta, \theta}^{\dagger} \hat{d}_{k; \epsilon, \zeta}^{\dagger} \hat{c}_{i; \alpha} \hat{c}_{j; \beta} \hat{c}_{k; \gamma} \hat{c}_{l; \delta}\) & \({ DD}:{ dd}\to{ EE}\) & \(T_{\alpha, \beta, \gamma, \delta}^{(\epsilon, \zeta), (\eta, \theta)} \Delta_{\gamma}^{\epsilon, \zeta} \Delta_{\delta}^{\eta, \theta}\)\\
\end{tabular}
\end{table*}

\begin{table*}[htbp]
\caption{\label{tab:DMtab} Monomer-dimer classes}
\centering
\begin{tabular}{llll}
class & operator & adjoint class & Spin term\\
\hline
\({ DM}:\varnothing \to \varnothing\) & \(\hat{c}_{j; \delta}^{\dagger} \hat{d}_{i; \epsilon, \zeta}^{\dagger} \hat{d}_{i; \alpha, \beta} \hat{c}_{j; \gamma}\) & self & \(C\left(T_{(\alpha, \beta), \gamma}^{\delta, (\epsilon, \zeta)}\Delta_{\alpha, \beta}^{\epsilon, \zeta}\right)\)\\
\({ DM}:{ m}\to{ E}\) & \(\hat{c}_{k; \delta}^{\dagger} \hat{d}_{i; \epsilon, \zeta}^{\dagger} \hat{d}_{i; \alpha, \beta} \hat{c}_{j; \gamma}\) & self & \(C\left(T_{(\alpha, \beta), \gamma}^{\delta, (\epsilon, \zeta)} \Delta_{\alpha, \beta}^{\epsilon, \zeta} \right)\)\\
\({ DM}:{ m}\to{ M}\) & \(\hat{d}_{k; \epsilon, \zeta}^{\dagger} \hat{d}_{i; \eta, \theta}^{\dagger} \hat{d}_{i; \alpha, \beta} \hat{c}_{j; \gamma} \hat{c}_{k; \delta}\) & \({ DD}:{ d}\to{ E}\) & \(T_{(\alpha, \beta), \gamma, \delta}^{(\epsilon, \zeta), (\eta, \theta)} \Delta_{\delta}^{\epsilon, \zeta}\Delta_{\alpha, \beta}^{\eta, \theta}\)\\
\({ DM}:{ d}\to{ E}\) & \(\hat{c}_{k; \delta}^{\dagger} \hat{c}_{j; \epsilon}^{\dagger} \hat{c}_{i; \zeta}^{\dagger} \hat{d}_{i; \alpha, \beta} \hat{c}_{j; \gamma}\) & \({ MM}:{ m}\to{ M}\) & \(T_{(\alpha, \beta), \gamma}^{\delta, \epsilon, \zeta} \Delta_{\zeta}^{\alpha, \beta}\)\\
\({ DM}:{ d}\to{ M}\) & \(\hat{d}_{k; \epsilon, \zeta}^{\dagger} \hat{c}_{j; \eta}^{\dagger} \hat{c}_{i; \theta}^{\dagger} \hat{d}_{i; \alpha, \beta} \hat{c}_{j; \gamma} \hat{c}_{k; \delta}\) & self & \(C\left(T_{(\alpha, \beta), \gamma, \delta}^{(\epsilon, \zeta), \eta, \theta} \Delta_{\delta}^{\epsilon,\zeta} \Delta_{\theta}^{\alpha, \beta}\right)\)\\
\({ DM}:{ dm}\to{ EE}\) & \(\hat{c}_{l; \epsilon}^{\dagger} \hat{c}_{k; \delta}^{\dagger} \hat{c}_{i; \zeta}^{\dagger} \hat{d}_{i; \alpha, \beta} \hat{c}_{j; \gamma}\) & \({ MM}:{ mm}\to{ EM}\) & \(T_{(\alpha, \beta), \gamma}^{\delta, \epsilon, \zeta} \Delta_{\zeta}^{\alpha, \beta}\)\\
\({ DM}:{ dm}\to{ E}^2\) & \(\hat{d}_{k; \delta, \epsilon}^{\dagger} \hat{c}_{i; \zeta}^{\dagger} \hat{d}_{i; \alpha, \beta} \hat{c}_{j; \gamma}\) & \({ D}:{ d}^2\to{ EM}\) & \(T_{(\alpha, \beta), \gamma}^{(\delta, \epsilon), \zeta} \Delta_{\zeta}^{\alpha, \beta}\)\\
\({ DM}:{ dm}\to{ EM}\) & \(\hat{c}_{l; \eta}^{\dagger} \hat{d}_{k; \epsilon, \zeta}^{\dagger} \hat{c}_{i; \theta}^{\dagger} \hat{d}_{i; \alpha, \beta} \hat{c}_{j; \gamma} \hat{c}_{k; \delta}\) & self & \(C\left(T_{(\alpha, \beta), \gamma, \delta}^{(\epsilon, \zeta), \eta, \theta} \Delta_{\delta}^{\epsilon, \zeta} \Delta_{\theta}^{\alpha, \beta}\right)\)\\
\({ DM}:{ dm}\to{ MM}\) & \(\hat{d}_{l; \theta, \iota}^{\dagger} \hat{d}_{k; \zeta, \eta}^{\dagger} \hat{c}_{i; \kappa}^{\dagger} \hat{d}_{i; \alpha, \beta} \hat{c}_{j; \gamma} \hat{c}_{k; \delta} \hat{c}_{l; \epsilon}\) & \({ DD}:{ dd}\to{ EM}\) & \(T_{(\alpha, \beta), \gamma, \delta, \epsilon}^{(\zeta, \eta), (\theta, \iota), \kappa} \Delta_{\kappa}^{\alpha, \beta} \Delta_{\delta}^{\zeta, \eta} \Delta_{\epsilon}^{\theta, \iota}\)\\
\end{tabular}
\end{table*}

\begin{table*}[htbp]
\caption{\label{tab:DDtab} dimer-dimer classes}
\centering
\begin{tabular}{llll}
class & operator & adjoint class & Spin term\\
\hline
\({ DD}:\varnothing\to\varnothing\) & \(\hat{d}_{j; \epsilon, \zeta}^{\dagger} \hat{d}_{i; \eta, \theta}^{\dagger} \hat{d}_{i; \alpha, \beta} \hat{d}_{j; \gamma, \delta}\) & self & \(C\left(T_{(\alpha, \beta), (\gamma, \delta)}^{(\epsilon, \zeta), (\eta, \theta)} \Delta_{\alpha, \beta}^{\eta, \theta} \Delta_{\gamma, \delta}^{\epsilon, \zeta}\right)\)\\
\({ DD}:{ d}\to{ E}\) & \(\hat{c}_{k, \epsilon}^{\dagger} \hat{c}_{j; \zeta}^{\dagger} \hat{d}_{i; \eta, \theta}^{\dagger} \hat{d}_{i; \alpha, \beta} \hat{d}_{j; \gamma, \delta}\) & \({ DM}:{ m}\to{ M}\) & \(T_{(\alpha, \beta), (\gamma, \delta)}^{\epsilon, \zeta, (\eta, \theta)} \Delta_{\zeta}^{\gamma, \delta}\Delta_{\alpha, \beta}^{\eta, \theta}\)\\
\({ DD}:{ d}\to{ M}\) & \(\hat{d}_{k; \zeta, \eta}^{\dagger} \hat{c}_{j, \theta}^{\dagger} \hat{d}_{i; \iota, \kappa}^{\dagger} \hat{d}_{i; \alpha, \beta} \hat{d}_{j; \gamma, \delta} \hat{c}_{k; \epsilon}\) & self & \(C\left(T_{(\alpha, \beta), (\gamma, \delta), \epsilon}^{(\zeta, \eta), \theta, (\iota, \kappa)}\Delta_{\epsilon}^{\zeta, \eta} \Delta_{\theta}^{\gamma, \delta} \Delta_{\alpha, \beta}^{\iota, \kappa}\right) \)\\
\({ DD}:{ dd}\to{ EE}\) & \(\hat{c}_{l; \zeta}^{\dagger}\hat{c}_{k; \epsilon}^{\dagger}\hat{c}_{j; \eta}^{\dagger}\hat{c}_{i; \theta}^{\dagger} \hat{d}_{i; \alpha, \beta} \hat{d}_{j; \gamma, \delta}\) & \({ MM}:{ mm}\to{ MM}\) & \(T_{(\alpha, \beta), (\gamma, \delta)}^{\epsilon, \zeta, \eta, \theta} \Delta_{\theta}^{\alpha, \beta} \Delta_{\eta}^{\gamma, \delta}\)\\
\({ DD}:{ dd}\to{ E}^2\) & \(\hat{d}_{k; \epsilon, \zeta}^{\dagger} \hat{c}_{j; \eta}^{\dagger}\hat{c}_{i; \theta}^{\dagger} \hat{d}_{i; \alpha, \beta} \hat{d}_{j; \gamma, \delta}\) & \({ D}:{ d}^2\to{ MM}\) & \(T_{(\alpha, \beta), (\gamma, \delta)}^{(\epsilon, \zeta), \eta, \theta}\Delta_{\theta}^{\alpha, \beta} \Delta_{\eta}^{\gamma, \delta}\)\\
\({ DD}:{ dd}\to{ EM}\) & \(\hat{c}_{l; \theta}^{\dagger} \hat{d}_{k; \zeta, \eta}^{\dagger} \hat{c}_{j; \iota}^{\dagger} \hat{c}_{i; \kappa}^{\dagger} \hat{d}_{i; \alpha, \beta} \hat{d}_{j; \gamma, \delta} \hat{c}_{k; \epsilon}\) & \({ DM}:{ dm}\to{ MM}\) & \(T_{(\alpha, \beta), (\gamma, \delta), \epsilon}^{(\zeta, \eta), \theta, \iota, \kappa} \Delta_{\kappa}^{\alpha,\beta} \Delta_{\iota}^{\gamma, \delta} \Delta_{\epsilon}^{\zeta, \eta}\)\\
\({ DD}:{ dd}\to{ MM}\) & \(\hat{d}_{l; \iota, \kappa}^{\dagger} \hat{d}_{k; \eta, \theta}^{\dagger} \hat{c}_{j; \mu}^{\dagger} \hat{c}_{i; \lambda}^{\dagger}\hat{d}_{i; \alpha, \beta} \hat{d}_{j; \gamma, \delta} \hat{c}_{k; \epsilon} \hat{c}_{l; \zeta}\) & self & \(C\left(T_{(\alpha, \beta), (\gamma, \delta), \epsilon, \zeta}^{(\eta, \theta), (\iota, \kappa), \lambda, \mu} \Delta_{\lambda}^{\alpha, \beta} \Delta_{\mu}^{\gamma, \delta} \Delta_{\epsilon}^{\eta, \theta} \Delta_{\zeta}^{\iota, \kappa}\right)\)\\
\end{tabular}
\end{table*}

\subsubsection{Two-body Hamiltonian, $\hat{H}_2$: Hubbard form}
In principle, an exact projection of the {\it ab initio} Hamiltonian on the monomer-dimer space generates all of the terms in the Tables~\ref{tab:Dtab}, \ref{tab:MMtab}, \ref{tab:DMtab}, and \ref{tab:DDtab}. However, to proceed we make the Hubbard approximation by dropping all terms corresponding to the motion of particles among the sites as a result of the interaction. This is justified in the case that the lattice is deep enough, since we consider density-density interactions and the Wannier functions are well-localized in their corresponding sites. The density-density character of interactions excludes particle-transfer terms at the level of the continuum model, while the localization of the Wannier functions ensures that density overlap integrals result to negligible off-diagonal contributions in the site basis. With this assumption only the interaction classes ${ D}:\varnothing \to \varnothing$, ${ MM}:\varnothing \to \varnothing$, ${ DM}:\varnothing \to \varnothing$ and ${ DD}:\varnothing \to \varnothing$ need to be considered. By explicitly calculating the corresponding spin-terms we obtain the Hubbard form of the interaction Hamiltonian
\begin{widetext}
\begin{align}
  \hat{H}_2^{\rm Hubbard} =& V_{ D} \sum_{i} \sum_{\alpha < \beta} \hat{d}_{i;\alpha, \beta}^{\dagger} \hat{d}_{i;\alpha, \beta} \notag \\
  &+ \sum_{i \neq j} \frac12 V_{ MM}^{\rm dir}(i,j) \sum_{\alpha,\beta} \hat{c}_{j;\beta}^{\dagger} \hat{c}_{i;\alpha}^{\dagger}\hat{c}_{i;\alpha}\hat{c}_{j;\beta} \notag \\
  &+ \sum_{i \neq j} \frac12 V_{ MM}^{\rm exc}(i,j) \sum_{\alpha,\beta} \hat{c}_{j;\alpha}^{\dagger} \hat{c}_{i;\beta}^{\dagger}\hat{c}_{i;\alpha}\hat{c}_{j;\beta} \notag \\
  &+ \sum_{i \neq j} V_{ DM}^{\rm dir}(i,j) \sum_{\alpha<\beta, \gamma} \hat{c}_{j;\gamma}^{\dagger} \hat{d}_{i;\alpha, \beta}^{\dagger} \hat{d}_{i;\alpha, \beta}\hat{c}_{j;\gamma} \notag \\
  &+ \sum_{i \neq j} V_{ DM}^{\rm exc}(i,j) \sum_{\alpha<\beta, \gamma} \left( \hat{c}_{j;\beta}^{\dagger} \hat{d}_{i;\alpha, \gamma}^{\dagger} + \hat{c}_{j;\alpha}^{\dagger} \hat{d}_{i;\gamma, \beta}^{\dagger} \right) \hat{d}_{i;\alpha, \beta}\hat{c}_{j;\gamma} \notag \\
  &+ \sum_{i \neq j} \frac12 V_{ DD}^{\rm dir}(i,j) \sum_{\alpha<\beta, \gamma < \delta} \hat{d}_{j;\gamma, \delta}^{\dagger} \hat{d}_{i;\alpha, \beta}^{\dagger} \hat{d}_{i;\alpha, \beta}\hat{d}_{j;\gamma, \delta} \notag \\
  &+ \sum_{i \neq j} \frac12 V_{ DD}^{\rm sgl}(i,j) \sum_{\alpha < \beta, \gamma < \delta, \epsilon < \zeta, \eta < \theta}
    \sigma_{\alpha, \beta, \gamma, \delta}^{\epsilon, \zeta, \eta, \theta} \hat{d}_{j;\eta, \theta}^{\dagger} \hat{d}_{i;\epsilon, \zeta}^{\dagger} \hat{d}_{i;\alpha, \beta}\hat{d}_{j;\gamma, \delta},
\end{align}
where $\sigma_{\alpha_1, \alpha_{2}, \alpha_3, \alpha_4}^{\beta_1, \beta_2, \beta_3, \beta_4} = \delta_{\alpha_1, \alpha_{2}, \alpha_3, \alpha_4}^{\beta_1, \beta_2, \beta_3, \beta_4} - \delta_{\alpha_1, \alpha_{2}}^{\beta_1, \beta_2} \delta_{\alpha_3, \alpha_{4}}^{\beta_3, \beta_4}- \delta_{\alpha_1, \alpha_{2}}^{\beta_3, \beta_4} \delta_{\alpha_3, \alpha_{4}}^{\beta_1, \beta_2}$ with
\end{widetext}
\begin{equation}
  \delta_{\alpha_1, \dots, \alpha_{N_{\sigma}}}^{\beta_1, \dots, \beta_{N_{\sigma}}} = {\rm det}\left(
\begin{array}{cc c c}
 \delta_{\alpha_1}^{\beta_1}& \delta_{\alpha_1}^{\beta_2}& \dots & \delta_{\alpha_1}^{\beta_{N_{\sigma}}}\\
 \delta_{\alpha_2}^{\beta_1}& \delta_{\alpha_2}^{\beta_2}& \dots & \delta_{\alpha_2}^{\beta_{N_{\sigma}}}\\
  \vdots & \vdots & & \vdots \\
 \delta_{\alpha_{N_{\sigma}}}^{\beta_1}& \delta_{\alpha_{N_{\sigma}}}^{\beta_2}& \dots & \delta_{\alpha_{N_{\sigma}}}^{\beta_{N_{\sigma}}}\\
\end{array} \right).
\end{equation}
Notice that the term proportional to $V^{\rm exc}_{DM}(i, j)$ also contains dimer operators with inverted order of spin-indices, in this case the use of $\hat{d}_{i;\beta, \alpha} = -\hat{d}_{i;\alpha, \beta}$ is implied to cast the spin-indices in ascending order.
The interaction couplings can be evaluated as
\begin{equation}
\begin{split}
V_D =&
\langle 0 |
\hat{d}_{i;1,2}
\hat{H}_2
\hat{d}_{i;1,2}^{\dagger}
|0\rangle, \\
    V_{ MM}^{\rm dir}(i,j) =& \langle 0 | \hat{c}_{j; 2}\hat{c}_{i; 1} \hat{H}_2 \hat{c}_{i; 1}^{\dagger}\hat{c}_{j; 2}^{\dagger} | 0 \rangle, \\
    V_{ MM}^{\rm exc}(i,j) =& \langle 0 | \hat{c}_{j; 1}\hat{c}_{i; 1} \hat{H}_2 \hat{c}_{i; 1}^{\dagger}\hat{c}_{j; 1}^{\dagger} | 0 \rangle \\
    &- \langle 0 | \hat{c}_{j; 2}\hat{c}_{i; 1} \hat{H}_2 \hat{c}_{i; 1}^{\dagger}\hat{c}_{j; 2}^{\dagger} | 0 \rangle, \\
    V_{ DM}^{\rm dir}(i, j) =& \langle 0 | \hat{d}_{i; 1, 2} \hat{c}_{j; 3} \hat{H}_2 \hat{c}_{j; 3}^{\dagger}\hat{d}_{i; 1, 2}^{\dagger} | 0 \rangle, \\
    V_{ DM}^{\rm exc}(i, j) =& \langle 0 | \hat{d}_{i; 1, 2} \hat{c}_{j; 1} \hat{H}_2 \hat{c}_{j; 1}^{\dagger}\hat{d}_{i; 1, 2}^{\dagger} | 0 \rangle \\
    &- \langle 0 | \hat{d}_{i, 1, 2} \hat{c}_{j; 3} \hat{H}_2 \hat{c}_{j; 3}^{\dagger}\hat{d}_{i; 1, 2}^{\dagger} | 0 \rangle, \\
    V_{ DD}^{\rm dir}(i, j) =& \langle 0 | \hat{d}_{j; 3, 4} \hat{d}_{i; 1, 2} \hat{H}_2 \hat{d}_{i; 1, 2}^{\dagger}\hat{d}_{j; 3, 4}^{\dagger} | 0 \rangle, \\
    V_{ DD}^{\rm sgl}(i, j) =& \langle 0 | \hat{d}_{j; 3, 4} \hat{d}_{i; 1, 2} \hat{H}_2 \hat{d}_{i; 1, 2}^{\dagger}\hat{d}_{j; 3, 4}^{\dagger} | 0 \rangle \\
    & -\langle 0 | \hat{d}_{j; 2, 3} \hat{d}_{i; 1, 2} \hat{H}_2 \hat{d}_{i; 1, 2}^{\dagger}\hat{d}_{j; 2, 3}^{\dagger} | 0 \rangle  \\
    =& \frac{1}{2} \bigg( 
    \langle 0 | \hat{d}_{j; 3, 4} \hat{d}_{i; 1, 2} \hat{H}_2 \hat{d}_{i; 1, 2}^{\dagger}\hat{d}_{j; 3, 4}^{\dagger} | 0 \rangle \\
    &-\langle 0 | \hat{d}_{j; 1, 2} \hat{d}_{i; 1, 2} \hat{H}_2 \hat{d}_{i; 1, 2}^{\dagger}\hat{d}_{j; 1, 2}^{\dagger} | 0 \rangle \bigg),
\end{split}
\label{Hubbard_parameters}
\end{equation}
where operators with spin indices outside the relevant range $\alpha \in \{1,\dots,N\}$ are understood to annihilate the state and thus make the irrelevant for this $N$ terms equal to zero.
We found that the interaction coefficients associated with spin-exchange processes (${\rm exc}$ superscript) are typically much smaller than those of the corresponding spin-independent interactions (${\rm dir}$ superscript). For example, numerical tests yield $V_{ MM}^{\rm exc}/V_{ MM}^{\rm dir}\approx 2\%$ and  $V_{ DM}^{\rm exc}/V_{ DM}^{\rm dir} < 0.1\%$. We therefore retain only the ``dir'' terms in our model presented in the main text.
\subsection{Monomer Gaussian basis}
To simplify our derivations we develop a Gaussian model for the projection of the {\it ab initio} Hamiltonian in the lowest band of the optical lattice. The idea is that for deep lattices the potential a particle experiences is akin to a harmonic oscillator and thus we can simplify $V_{{\rm latt},\mu}(r_\mu) = V_{0,\mu} \sin^{2} k_{\rm L} r_{\mu} = V_{0,\mu} k_{\rm L}^2 r_{\mu}^2 + \mathcal{O}(r_{\mu}^4)$, $\mu \in \{x, y, z\}$. This allows us to approximate the states at each lattice site with the corresponding ground-state solution of the harmonic oscillator. We fix the frequency, $\omega_{{\rm ho}, \mu} = \frac{2 E_{\rm R}}{\hbar} \sqrt{\frac{V_{0, \mu}}{E_{\rm R}}}$, and length scale, $a_{{\rm ho},\mu} = \sqrt{\frac{\hbar}{m \omega_{{\rm ho}, \mu}}} = \frac{1}{k}\left( \frac{E_{\rm R}}{V_{0,\mu}} \right)^{1/4}$, according to the Taylor expansion of $V_{{\rm latt};\mu}(r_{\mu})$. The monomer solution is therefore a Gaussian.

Notice that this basis is not orthonormal since the states at different sites have overlap
\begin{equation}
\begin{split}
S_{j, j+l}&=\int_{-\infty}^{\infty} \mathrm{d}r_{\mu}~\psi_0^{*}(r_{\mu} -R_{j;\mu}) \psi_0(r_{\mu} -R_{j;\mu} - l a_{\rm latt}) \\
          &= \langle 0 | D\left( \frac{l \pi}{\sqrt{2} k_{\rm L} a_{{\rm ho},\mu}} \right) | 0 \rangle 
          = \exp \left(- \frac{1}{4} \frac{l^2 \pi^{2}}{k_{\rm L}^2 a_{{\rm ho},\mu}^2} \right),
\end{split}
\end{equation}
where $D(\cdot)$ denotes the displacement operator which is discussed in more detail later in this Supplementary Material.
The matrix elements of the single-particle Hamiltonian read
\begin{equation}
\begin{split}
  H_{j, j+l}=& \int_{-\infty}^{\infty} \mathrm{d}r_{\mu}~\psi_0^{*}(r_{\mu} -R_{j;\mu}) \hat{H}_0 \psi_0(r_{\mu} -R_{j;\mu} - l a_{\rm latt}) \\
            =& -\frac{V_{0;\mu}}{2} \exp \left(- \frac{1}{4} \frac{l^2 \pi^{2}}{k_{\rm L}^2 a_{{\rm ho},\mu}^2} \right) \\
              & \times \bigg[ \frac{1}{2} \frac{l^{2} \pi^{2}}{k_{\rm L}^4 a_{{\rm ho},\mu}^4} \frac{E_{\rm R}}{V_{0; \mu}}
                - \left(1 + \frac{E_{\rm R}}{V_{0;\mu}}\frac{1}{k_{\rm L}^2 a_{{\rm ho},\mu}^2} \right) \\
  &\hspace{0.5cm}+ (-1)^l \exp\left(- k_{\rm L}^2 a_{{\rm ho},\mu}^2 \right) \bigg].
\end{split}
\end{equation}
Using these matrix elements, one can obtain the tunnelling parameter $t_G$:
\begin{align}
t_G&=\frac{1}{8}e^{-\frac{\pi^2}{4 a_{\rm ho}^2k_{\rm L}^2}}\left[\frac{2(2+\lambda)}{ma_{\rm ho}^2}-\frac{\pi^2}{mk_{\rm L}^2a_{\rm ho}^4} \right. \notag \\
&+\left. 4V_0\left(2+\lambda^2-\lambda^2e^{-\frac{k_{\rm L}^2a_{\rm ho}^2}{\lambda}}\right)\right]
\end{align}
in the Gaussian model.

The value of tunneling parameter $t$ can be extracted from the exact-bandwidth obtained by the exact diagonalization of $V_{{\rm trap};\mu}(r_{\mu})$ in terms of Mathieu functions~\cite{Zwerger2003}
\begin{equation}\label{eq:tw}
\begin{split}
&t = \frac{E_{\rm R}}{4} \left[ b_{1}\left(\frac{V_0}{4 E_{\rm R}}\right) - a_{0}\left(\frac{V_0}{4 E_{\rm R}}\right) \right] \\
 &\approx \frac{4}{\sqrt{\pi}} \left(\frac{V_0}{E_{\rm R}}\right)^{3/4} e^{-2 \sqrt{\frac{V_0}{E_{\rm R}}}} \left( 1 - \frac{7}{16}\sqrt{\frac{E_{\rm R}}{V_0}} + \dots \right),
\end{split}
\end{equation}
where $a_{n-1}(q)$, $b_n(q)$, with $n = 1, 2, \dots$, are the Mathieu characteristic numbers.

We find that the hopping amplitude $t_G$ obtained from the bare Gaussian model agrees well with $t$ calculated from Eq.~(\ref{eq:tw}), particularly when $s$ is small, as shown in Fig.~\ref{fig:gam}(b) of the main text. 
We use $t_G$ to renormalize the other Hubbard parameters, as they are computed within the same level of approximation.

\subsection{Numerical method for off-site interactions and tunnelling parameters}
To numerically obtain the tunnelling parameters and the off-site interaction coefficients, we incorporate the displacement operator  
\begin{equation}
 D_j(\alpha) = \exp \left( \alpha \hat{a}^{\dagger}_j - \alpha^* \hat{a}_j \right)
\end{equation}
into the framework of harmonic oscillator brackets \cite{KAMUNTAVICIUS2001,Germanas2010}. Here, the  $\hat{a}_j$ and $\hat{a}^{\dagger}_j$ denote the annihilation and creation operators of the harmonic-oscillator mode associated with particle $j$: 
 \begin{equation} \label{eq:aj}
    \hat{a}_j = \frac{1}{\sqrt{2}} \left( \frac{\hat{r}_j}{a_{\rm ho}} + i \frac{\hat{p}_j a_{\rm ho}}{\hbar} \right), \;
    \hat{a}^{\dagger}_j = \frac{1}{\sqrt{2}} \left( \frac{\hat{r}_j}{a_{\rm ho}} - i \frac{\hat{p}_j a_{\rm ho}}{\hbar} \right),
\end{equation}
respectively, where $a_{\rm ho}=\sqrt{\hbar/m\omega_{\rm ho}}$ is the single-monomer oscillator length. They should not be confused with $\hat{c}^{\dagger}_{j;\alpha}$ and $\hat{c}_{j;\alpha}$, which create and annihilate a monomer at site $j$ and spin $\alpha$ in the Fock-space representation. One can rewrite the  $D_j(\alpha)$ as
\begin{equation}
\begin{split}
    D_j(\alpha) =& \exp \left( i \sqrt{2} {\rm Im} \alpha \frac{\hat{r}_j}{a_{\rm ho}} - i \sqrt{2} {\rm Re} \alpha \frac{a_{\rm ho} \hat{p}_j}{\hbar} \right), \\
\end{split}
\end{equation}
and as
\begin{align}
    D_j(\alpha) =& \exp \left( i \sqrt{2} {\rm Im} \alpha \frac{\hat{R}_{ij}}{a_{\rm ho}} - i \frac{{\rm Re} \alpha}{\sqrt{2}}  \frac{a_{\rm ho} \hat{P}_{ij}}{\hbar} \right) \\
    &\times \exp \left( i \frac{{\rm Im} \alpha}{\sqrt{2}} \frac{\hat{r}_{ij}}{a_{\rm ho}} - i \sqrt{2} {\rm Re} \alpha \frac{a_{\rm ho} \hat{p}_{ij}}{\hbar} \right),
\end{align}
in the representation of the centre of mass ($\hat{R}_{ij}=\frac{\hat{r}_i+\hat{r}_j}{2},\hat{P}_{ij}=\hat{p}_i+\hat{p}_j$) and relative ($\hat{r}_{ij}=\hat{r}_j-\hat{r}_i$, $\hat{p}_{ij}=\frac{\hat{p}_j-\hat{p}_i}{2}$) motion. The displacement operators $D^R_{ij}$ and $D^r_{ij}$ in the center of mass and relative frames, respectively, are defined by substituting the coordinate, momentum, and oscillator length appearing in Eq. (\ref{eq:aj}) by their correspondences in each frame. Such a definition gives
\begin{equation}
D_j(\alpha)=D^R_{ij}\left(\frac{\alpha}{\sqrt{2}}\right)D^r_{ij}\left(\frac{\alpha}{\sqrt{2}}\right),
\end{equation}
connecting the displacement operator in the single-monomer representation to that in center of mass and the relative representation.

 The transformation of the harmonic oscillator basis between the single-particle representation and the two-particle center-of-mass and relative-coordinate representation is carried out via
  \begin{align}
 &|H_{ij}\rangle|h_{ij}\rangle =\sum_{h_ih_j}[U]_{h_ih_j}^{H_{ij}h_{ij}}|h_i\rangle|h_j\rangle , \\
 &|h_i\rangle|h_j\rangle=\sum_{H_{ij}h_{ij}}[U^{-1}]_{H_{ij}h_{ij}}^{h_ih_j} |H_{ij}\rangle |h_{ij}\rangle,
 \end{align}
 where $|h_i=n_il_im_i\rangle$, $|h_{ij}=n_{r_{ij}}l_{r_{ij}}m_{r_{ij}}\rangle$ and $|H_{ij}=n_{R_{ij}}l_{R_{ij}}m_{R_{ij}}\rangle$ denote the harmonic oscillator basis of a single monomer, and that of a pair of monomers in center of mass and relative motion, respectively. The transformation matrix is given by
 \begin{align}
 [U]_{h_ih_j}^{Hh} &=\langle n_il_im_i;n_jl_jm_j|n_{R_{ij}}l_{R_{ij}}m_{R_{ij}};n_{r_{ij}}l_{r_{ij}}m_{r_{ij}}\rangle \notag \\
 &=\sum_{\Lambda = |l_i - l_j|}^{l_i + l_j} \sum_{M = -\Lambda}^{\Lambda} C^{\Lambda M}_{l_i, m_j; l_i, m_j} \sum_{m_i, m_j} C^{\Lambda, M}_{l_i, m_i; l_j, m_j} \notag \\ 
    &\times \sum_{\{n_i, n_j, l_i, l_j\}\in \mathcal{A}} \langle n_i l_i; n_j l_j; \Lambda | n_{R_{ij}} l_{R_{ij}}; n_{r_{ij}} l_{r_{ij}}; \Lambda \rangle_d, 
\end{align}
where the set of allowed indices satisfies the equations 
\begin{equation}
\begin{split}
    \mathcal{A} = \{&(n_i, n_j, l_i, l_j) \in \mathbb{N}^4: \\ 
    &2 n_i + l_i + 2 n_j + l_j = 2 n_{R_{ij}} + l_{R_{ij}} + 2 n_{r_{ij}} + l_{r_{ij}}, \\ &(-1)^{l_i + l_j} = (-1)^{l_{R_{ij}} + l_{r_{ij}}} \},
\end{split}
\end{equation}
and $C^{\Lambda M}_{l_i, m_j; l_i, m_j}$ is the Clebsch-Gordan coefficient. The harmonic oscillator bracket  $\langle n_i l_i; n_j l_j; \Lambda | n_{R_{ij}} l_{R_{ij}}; n_{r_{ij}} l_{r_{ij}}; \Lambda \rangle_d$ can be calculated by publicly available codes \cite{Germanas2010}
at $d=1$, which parametrizes the rotation from the lab to the relative frame
\begin{equation}
\left( \begin{array}{c}
     {\bm R_{ij}}  \\
     {\bm r_{ij}} 
\end{array} \right) = 
\left( \begin{array}{c c}
     \sqrt{\frac{d}{1+d}} & \sqrt{\frac{1}{1+d}} \\
     \sqrt{\frac{1}{1+d}} & -\sqrt{\frac{d}{1+d}}
\end{array} \right) 
\left( \begin{array}{c}
     {\bm r}_i  \\
     {\bm r}_j 
\end{array} \right) 
\end{equation}
\subsubsection{The off-site interactions}

By neglecting the interaction terms associated with spin-exchanging processes, only three types of off-site interaction parameters $V^{\rm dir}_{MM}(i,j)\rightarrow V^{\sigma}_{MM}$,  $V^{\rm dir}_{DM}(i,j)\rightarrow V^{\sigma}_{DM}$ and $V^{\rm dir}_{DD}(i,j)\rightarrow V^{\sigma}_{DD}$ remain in our Hubbard model. The superscript $\sigma = n x + m y$, $n \in \mathbb{Z}$ indicates relative displacement of the two involved sites in integer units $n, m$ of sites. Following Eq.~\eqref{Hubbard_parameters}, these parameters correspond to the matrix elements of the two-body interaction Hamiltonian 
\begin{equation}
\begin{split}
\hat{H}_2 &= \frac{1}{2} \sum_{\alpha} \hat{H}_{\alpha \alpha} + \sum_{\alpha < \beta} \hat{H}_{\alpha \beta},\\
\hat{H}_{\alpha \beta} &= \int {\rm d}^3r_1{\rm d}^3r_2\,
 \hat{\Psi}_{\alpha}^{\dagger}({\bm r}_1)\hat{\Psi}_{\beta}^{\dagger}({\bm r}_2)\\
&\hspace{65pt} \times V_{\rm sh}({\bm r}_2 - {\bm r}_1)\hat{\Psi}_{\beta}({\bm r}_2)\hat{\Psi}_{\alpha}({\bm r}_1),
\end{split}
\end{equation}
where $\hat{\Psi}_{\alpha}({\bm r})$ is the spin-$\alpha$ field operator. The matrix elements involve the monomer-monomer $|\psi_{MM}\rangle \equiv \hat{c}_{i; 1}^{\dagger}\hat{c}_{j; 2}^{\dagger} | 0 \rangle$, monomer-dimer $|\psi_{DM}\rangle\equiv \hat{d}_{i; 1, 2}^{\dagger}\hat{c}_{j; 3}^{\dagger} | 0 \rangle$ and dimer-dimer $|\psi_{DD}\rangle=\hat{d}_{i; 1, 2}^{\dagger}\hat{d}_{j; 3, 4}^{\dagger} | 0 \rangle$ states.
The monomer and dimer creation operators are defined as
\begin{equation}
\begin{split}
    \hat{c}_{i; \alpha}^{\dagger} &= \int {\rm d}^3r\, \psi_{\rm m}({\bm r}-{\bm R}_i) \hat{\Psi}^{\dagger}_{\alpha}({\bm r})\\
    \hat{d}_{i; \alpha, \beta}^{\dagger} &= \int {\rm d}^3r_1 {\rm d}^3r_2\, \psi_{\rm CM}\left(\frac{{\bm r}_1+{\bm r}_2}{2}-{\bm R}_i\right) \\
    &\hspace{62pt}\times \psi_{\rm d}({\bm r}_2-{\bm r}_1) \hat{\Psi}^{\dagger}_{\alpha}({\bm r}_1)\hat{\Psi}^{\dagger}_{\beta}({\bm r}_2),
\end{split}
\end{equation}
where ${\bm R}_i$ is the site position relative to the site $i = 0$ with ${\bm R}_0 = {\bm 0}$ and $\psi_{\rm m}({\bm r})$, $\psi_{\rm CM}({\bm r})$, $\psi_{\rm d}({\bm r})$ correspond to the monomer, dimer center-of-mass and dimer relative wavefunctions respectively.

Since the spin-algebra is already accounted for in Eq.~\eqref{Hubbard_parameters}, we simplify our description by henceforth evaluating the involved spatial matrix elements within the first-quantization picture. After imposing the proper (anti)symmetrization required by the spin state, we consider particle-labeled coordinate operators $\hat{\bm r}_p$, with $p = \{1, 2, 3, 4\}$ the particle index. The corresponding ket states are defined via their wavefunctions, such that $\langle {\bm r}_p|\psi_p\rangle \equiv \psi_{\rm m}({\bm r}_p)$ for monomers and $\langle {\bm r}_p, {\bm r}_{p'}|\psi_{p, p'}\rangle = \psi_{\rm CM}(\frac{{\bm r}_p+{\bm r}_{p'}}{2}) \psi_{\rm d}({\bm r}_{p'} - {\bm r}_p)$ for dimers.
Within this description, states localized in different sites are represented by the action of the displacement operator that moves the particle $p$ to the desired site $j$. For instance, if $p$ corresponds to the spin-$\alpha$ particle in $\hat{c}^{\dagger}_{j; \alpha} |0 \rangle$, its corresponding first-quantization state is $|\psi_{p \to j} \rangle = \hat{D}_{p}(\frac{{\bm R}_j}{\sqrt{2} a_{\rm ho}}) | \psi_p \rangle$, similarly for a dimer composed of $p$, $p'$ the corresponding first-quantization state reads $|\psi_{p p' \to j} \rangle = \hat{D}_{p}(\frac{{\bm R}_j}{\sqrt{2} a_{\rm ho}}) \hat{D}_{p'}(\frac{{\bm R}_j}{\sqrt{2} a_{\rm ho}})| \psi_{p p'} \rangle$.

This enables a general approach to calculate the off-site interaction parameters: expand the corresponding wavefunction in the product state of single-monomer harmonic oscillator basis $\Pi_{i=1}^N\otimes|h_i\rangle$ and then transfer the subspace of interacting pair $(i,j)$ into two-monomer centre of mass and relative representation to evaluate the interaction matrix element. For instance, the monomer-monomer interaction $V_{MM}^{\sigma}$ is given by
\begin{align}
V_{MM}^{\sigma}&=\langle \psi_1\psi_2|D^{\dagger}_2(\alpha_\sigma)V_{\rm sh}(\bm{r}_{2}-\bm{r}_{1})D_2(\alpha_\sigma)|\psi_1 \psi_2\rangle \notag \\
&=\langle \psi_1\psi_2|D^{r\dagger}_{12}(\alpha_\sigma/\sqrt{2})V_{\rm sh}(\bm{r}_{12})D^r_{12}(\alpha_\sigma/\sqrt{2})|\psi_1 \psi_2\rangle \notag \\
&=\sum_{h_1h_2;h'_1h'_2}\sum_{H_{12}h_{12};H'_{12}h'_{12}}c^*_{h_1}c^*_{h_2}c_{h'_1}c_{h'_2}[U]_{h_1h_2}^{H_{12}h_{12}} \notag \\
&\times [U^{-1}]_{H'_{12}h'_{12}}^{h'_{1}h'_{2}}\delta_{H_{12}H'_{12}}\langle h_{12}|V_{\rm sh}^D(\bm{r}_{12};\alpha_\sigma/\sqrt{2})|h'_{12}\rangle,
\end{align}
where $V_{\rm sh}^D(\bm{r}_{12};\alpha_\sigma)=D^{r\dagger}_{12}(\alpha_\sigma)V_{\rm sh}(\bm{r}_{12})D^r_{12} (\alpha_\sigma)$ denote the displaced interaction and $c_{h_i}=\langle h_i|\psi_i\rangle$ is the component of single-monomer wavefunction in harmonic oscillator mode $|h_i=n_il_im_i\rangle$. The single monomer state $|\psi_i\rangle$ is the ground state of the anisotropic harmonic trap. Here, $\sigma=x,y,x+y$ or $x-y$ describes the displacement in $x,y,x+y$ or $x-y$ direction, respectively. The corresponding $\alpha^\sigma$ are listed in Table \ref{tab:alpha}.
\begin{table}[]
    \centering
    \caption{The parameters of displacement with $\tilde{\alpha}=a_{\rm latt}/(\sqrt{2}a_{\rm ho})$.}
    \begin{tabular}{ccccc}
    \hline
    \hline
         $\sigma$ & $x$ &$y$&$x+y$&$x-y$\\
        $\alpha^{\sigma}$ &$\tilde{\alpha}\hat{e}_x$ &$\tilde{\alpha}\hat{e}_y$&$\tilde{\alpha}(\hat{e}_x+\hat{e}_y)$& $\tilde{\alpha}(\hat{e}_x-\hat{e}_y)$\\
    \hline
    \end{tabular}
    \label{tab:alpha}
\end{table}
Similarly, the dimer-monomer $V_{DM}^{\sigma}$ and dimer-dimer $V_{DD}^{\sigma}$ interactions are given by
 \begin{align} \label{vdm}
 V_{DM}^{\rm \sigma}&=\langle \psi_{12}\psi_3|D^{\dagger}_3(\alpha_\sigma)[V_{\rm sh}(\bm{r}_{13})+V_{\rm sh}(\bm{r}_{23})]D_3(\alpha_\sigma)|\psi_{12}\psi_{3}\rangle \notag \\
 &=2\langle \psi_{12}\psi_3|D^{r\dagger}_{23}(\alpha_\sigma/\sqrt{2})V_{\rm sh}(\bm{r}_{23})D^r_{23}(\alpha_\sigma/\sqrt{2})|\psi_{12}\psi_{3}\rangle \notag \\
 &=2\sum_{H_{12}h_{12}h_3}\sum_{H'_{12}h'_{12}h'_3}\sum_{h_1,h_2;h'_1h'_2}\sum_{H_{23},h_{23}}\sum_{H'_{23}h'_{23}}c^*_{H_{12}}c^*_{h_{12}} \notag \\
 &\times c^*_{h_3}c_{H'_{12}}c_{h'_{12}}c_{h'_3}[U^{-1}]_{H_{12}h_{12}}^{h_{1}h_{2}}[U]_{h'_1h'_2}^{H'_{12}h'_{12}}[U]^{H_{23}h_{23}}_{h_{2}h_{3}} \notag \\
 &\times [U^{-1}]^{h'_2h'_3}_{H'_{23}h'_{23}}\delta_{h_1h'_1}\delta_{H_{23}H'_{23}} \langle h_{23}|V_{\rm sh}^D(\bm{r}_{23};\alpha_\sigma/\sqrt{2})|h'_{23}\rangle
 \end{align}
 and
  \begin{align} \label{vdm}
 V_{DD}^{\rm \sigma}&=\langle \psi_{12}\psi_{34}|D^{\dagger}_3(\alpha_\sigma)D^{\dagger}_4(\alpha_\sigma)[V_{\rm sh}(\bm{r}_{13})+V_{\rm sh}(\bm{r}_{23}) \notag \\
 &+V_{\rm sh}(\bm{r}_{14})+V_{\rm sh}(\bm{r}_{24})]D_4(\alpha_\sigma)D_3(\alpha_\sigma)|\psi_{12}\psi_{34}\rangle \notag \\
 &=4\langle \psi_{12}\psi_{34}|D^{r\dagger}_{23}(\alpha_\sigma/\sqrt{2})V_{\rm sh}(\bm{r}_{23})D_{23}(\alpha_\sigma/\sqrt{2})|\psi_{12}\psi_{34}\rangle \notag \\
 &=4\sum_{H_{12}h_{12}H_{34}h_{34}}\sum_{H'_{12}h'_{12}H'_{34}h'_{34}}\sum_{h_1,h_2h_3h_4;h'_1h'_2h'_3h'_4} \notag \\
 &\times \sum_{H_{23},h_{23}}\sum_{H'_{23}h'_{23}}c^*_{H_{12}}c^*_{h_{12}}c^*_{H_{34}} c^*_{h_{34}}c_{H'_{12}}c_{h'_{12}}c_{H'_{34}}c_{h'_{34}} \notag \\
  &\times [U^{-1}]_{H_{12}h_{12}}^{h_{1}h_{2}}
[U]_{h'_1h'_2}^{H'_{12}h'_{12}}[U^{-1}]_{H_{34}h_{34}}^{h_{3}h_{4}}
[U]_{h'_3h'_4}^{H'_{34}h'_{34}} \notag \\
&\times [U]^{H_{23}h_{23}}_{h_{2}h_{3}}[U^{-1}]^{h'_2h'_3}_{H'_{23}h'_{23}}\delta_{h_1h'_1}\delta_{h_4h'_4}\delta_{H_{23}H'_{23}} \notag \\
&\times \langle h_{23}|V_{\rm sh}^D(\bm{r}_{23};\alpha_\sigma/\sqrt{2})|h'_{23}\rangle,
 \end{align}
 where the factors of 2 and 4 reflect the number of interacting monomer–monomer pairs. By symmetry, all pairs related by monomer exchange within a dimer contribute equally to $V_{DM}^{\sigma}$ or $V_{DD}^{\sigma}$. The coefficient $c_{H_{ij}}=\langle H_{ij}|\psi^R_{ij}\rangle$ and $c_{h_{ij}}=\langle h_{ij}|\psi^r_{ij}\rangle$ denotes the $|H_{ij}\rangle$ and $|h_{ij}\rangle$ component of dimer wavefunction $|\psi_{ij}\rangle=|\psi_{ij}^{R}\rangle|\psi_{ij}^r\rangle$ in center-of-mass and relative degree of freedom, respectively. The center-of-mass dimer state $|\psi_{ij}^{R}\rangle$ is the ground mode of the anisotropic harmonic trap, while the relative dimer state $|\psi_{ij}^{r}\rangle$ is obtained from the DVR method introduced in the last section, taking the actual shielding potential into account.

 It remains to evaluate the matrix element of the displaced interaction operator $V_{\rm sh}^D(\bm{r};\alpha_\sigma)=D^{r\dagger}(\alpha_\sigma)V_{\rm sh}(\bm{r})D^r (\alpha_\sigma)$, where we have dropped the particle indices for notational simplicity, as the matrix elements do not depend on the particular monomer pair considered. This can be done by calculating the matrix elements of the displacement operator and the shielding interaction separately:
 \begin{align}
 \left[V_{\rm sh}^D(\alpha_{\sigma})\right]_{n_rl_rm_r}^{n'_rl'_rm'_r}&=\sum_{n_r^{''}n_r^{''}n_r^{''}}\sum_{n_r^{'''}l_r^{'''}m_r^{'''}}[D^{-1}(\alpha_{\sigma})]_{n_rl_rm_r}^{n_r^{''}n_r^{''}n_r^{''}} \notag \\
 &\times [V_{\rm sh}]_{n_r^{''}n_r^{''}n_r^{''}}^{n_r^{'''}l_r^{'''}m_r^{'''}}[D(\alpha_{\sigma})]_{n_r^{'''}l_r^{'''}m_r^{'''}}^{n'_rl'_rm'_r}.
 \end{align}
 Here, the superscript '$r$' in the displacement operator matrix is further omitted because $[D(\alpha_{\sigma})]$ are the same at given $\alpha_{\alpha}$ for both single particle motion and center of mass and relative motions of the dimer. We then transfer into the Cartesian representation to obtain $[D(\alpha_\sigma)]$:
 \begin{align}
 [D(\alpha_\sigma)]_{nlm}^{n'l'm'}&=\sum_{n_xn_yn_z}\sum_{n'_xn'_yn'_z}[C^{-1}]_{nlm}^{n_xn_yn_z}[C]_{n'_xn'_yn'_z}^{n'lm'} \notag \\
 &\times D^{\rm 1d}_{n_xn'_x}(\alpha_\sigma^x) D^{\rm 1d}_{n_yn'_y}(\alpha_\sigma^y) D^{\rm 1d}_{n_zn'_z},
 \end{align}
 where $\alpha^{\mu}_{\sigma}=\hat{e}_{\mu} \cdot \alpha_{\sigma}$ for $\mu=x,y,z$. The transformation matrix between the Cartesian and spherical bases $C_{n_xn_ynz}^{nlm}=\langle n_x n_yn_z|nlm\rangle$ can be derived using the SO(3) algebra and the one-dimensional displacement operator matrix $D^{\rm 1d}_{n_{\mu}n'_{\mu}}(\alpha^{\mu}_{\sigma})$ is given analytically \cite{Becker2026}.
\begin{figure}
    \centering
    \includegraphics[width=1.0 \linewidth]{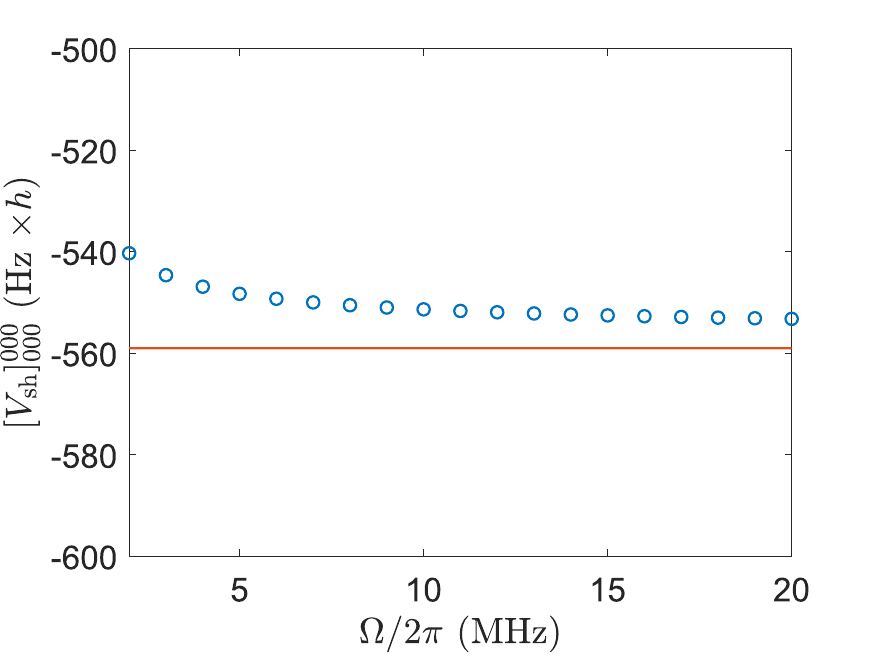}
    \caption{Comparison of $[V_{\rm sh}]_{000}^{000}$ calculated from actual numerical shielding potential (symbols) and from long-range tail approximation, i.e., $V_{\rm sh}(\bm{r})\approx C_3(3\cos^2\theta-1)/r^3$. The lattice parameters are chosen as $a_{\rm latt}=550$ nm, $s=5.1$ with $V_{0,z}=4V_{0,xy}$. }
    \label{fig:Vapprox}
\end{figure}

We calculate the shielding interaction matrix $[V_{\rm sh}]$ by numerical integration. We found that $V_{\rm sh}(\bm{r})\approx C_3(3\cos^2\theta-1)/r^3$ is a very good approximation in this calculation, see Fig. \ref{fig:Vapprox} for the comparison for the element $[V_{\rm sh}]_{000}^{000}$. The approximation slightly overestimates the magnitude of the interaction matrix element, with a deviation of less than $3\%$, for instance, for the tested $[V_{\rm sh}]_{000}^{000}$. This result indicates that the off-site interaction is predominantly determined by the long-range tail of the shielding potential. Therefore, throughout this work, we evaluate the off-site interaction parameters using the above approximation, which provides improved numerical stability. 

In the above analysis of $[V^{D}_{\rm sh}]$, we assume that the propagation direction of the microwave field is fixed along the $z$-axis of the lattice frame $(x,y,z)$. In the proposed rotation scheme, however, we introduce a microwave-field-fixed reference frame, $(x',y',z')$, which is related to the lattice frame by the Euler rotation $(\pi/4,\beta,-\pi/4)$. The $[V^{D}_{\rm sh}]$ will then be calculated in this reference frame and transferred back to the lattice frame via the Wigner D matrix. In such an approach, one should note that the argument $\alpha_\sigma$ of the displacement operator also needs to take the value in the new frame. For the subsequent analysis of the tunneling parameters, however, no frame rotation is required, and $\alpha_{\sigma}$ is always taken in the lattice frame, i.e., according to the values listed in Table~\ref{tab:alpha}. 
\subsubsection{The tunneling parameters}
We calculate the tunneling parameters $T$ and $T'$ by expanding the wavefunctions in the Cartesian harmonic basis,   $|h^{\rm C}_1=n_{x_1}n_{y_1}n_{z_1}\rangle|h^{\rm C}_2=n_{x_2}n_{y_2}n_{z_2}\rangle$ and $|h^{\rm C}_1=n_{x_1}n_{y_1}n_{z_1}\rangle|h^{\rm C}_2=n_{x_2}n_{y_2}n_{z_2}\rangle|h^{\rm C}_3=n_{x_3}n_{y_3}n_{z_3}\rangle$, respectively. Here, $|n_z\rangle$ is considered to be the harmonic oscillation mode in the $z$-direction associated with the same oscillation frequency $\omega_{\rm ho}$ as those in $x$ and $y$ directions.    
For tunneling along $x$, $T$ is given by:
\begin{align} \label{eq:Th}
T&=\langle\psi_{12}|[H_{\rm m}(\bm{r}_1)+H_{\rm m}(r_2)]D_2(\alpha_x)|\psi_1\psi_2\rangle \notag \\
&=\sum_{H_{12}h_{12}}\sum_{h_1h_2}\sum_{h'_1h'_2}\sum_{h^{\rm C}_1h^{\rm C}_2}\sum_{h^{\rm C'}_1h^{\rm C'}_2}c^*_{H_{12}}c^*_{h_{12}}c_{h_1}c_{h_2} \notag \\
&\times [U^{-1}]_{H_{12}h_{12}}^{h_{1}h_{2}}[C^{-1}]_{h_1}^{h^{\rm C}_1}[C^{-1}]_{h_2}^{h^{\rm C}_2}[C]_{h^{\rm C'}_1}^{h'_1}[C]_{h^{\rm C'}_2}^{h'_2} \notag \\
&\times\langle h^{\rm C}_1|\langle h^{\rm C}_2|[H_{\rm m}(\bm{r}_1)+H_{\rm m}(r_2)]D_2(\alpha_x)|h^{\rm C'}_1\rangle|h^{\rm C'}_2\rangle.
\end{align}
Here the single monomer Hamiltonian reads
\begin{align}
H_{\rm m}(\bm{r})&=\sum_{\mu=x,y,z}-\frac{\hbar^2}{2m}\nabla_{u}^2+V_{0,\mu}\sin^2(ku) \notag \\
&=\sum_{\mu=x,y,z}K_{\mu}+V_{0,\mu}\left[\frac{1}{2}\right.
-\left.\frac{1}{4}\sum_{s=\pm}D^{\rm 1d}_{\mu}(s\tilde{\alpha}_p)\right]
\end{align}
with $\tilde{\alpha}_p=i\sqrt{2}\pi a_{\rm ho}/a_{\rm latt}$, giving the matrix element $\langle h^{\rm C}_1|\langle h^{\rm C}_2|[H_{\rm m}(\bm{r}_1)+H_{\rm m}(r_2)]D_2(\alpha_x)|h^{\rm C'}_1\rangle|h^{\rm C'}_2\rangle$ as:
\begin{align}
&\langle h^{\rm C}_1|\langle h^{\rm C}_2|[H_{\rm m}(\bm{r}_1)+H_{\rm m}(r_2)]D_2(\alpha_x)|h^{\rm C'}_1\rangle|h^{\rm C'}_2\rangle \notag \\
&=\left(\sum_{n''_{x_2}}[K]_{n_{x_2},n''_{x_2}}[D^{\rm 1d}(\tilde{\alpha})]_{n''_{x_2},n'_{x_2}} \right. \notag \\
&+(2+\lambda^2)V_0*[D^{\rm 1d}(\tilde{\alpha})]_{n_{x_2},n'_{x_2}}  \notag \\
&-\left. \frac{V_0}{4}\sum_{n''_{x_2}}[\bar{D}^{\rm 1d}(\tilde{\alpha}_p)]_{n_{x_{2}},n''_{x_2}}[D^{\rm 1d}(\tilde{\alpha})]_{n''_{x_2},n'_{x_2}}\right) \notag \\
&\times \delta_{h^{\rm C}_1h^{\rm C}_2 \setminus n_{x_2},h^{\rm C'}_1h^{\rm C'}_2 \setminus n_{x'_2}} \notag \\
&+\sum_{\mu=x,y,i=1,2 \setminus x_2} \left([K]_{n_{\mu_{i}},n'_{\mu_i}}-\frac{V_0}{4}[\bar{D}^{\rm 1d}(\tilde{\alpha}_p)]_{n_{\mu_{i}},n'_{\mu_i}}\right) \notag \\
&\times [D^{\rm 1d}(\tilde{\alpha})]_{n_{x_2},n'_{x_2}}\delta_{h^{\rm C}_1h^{\rm C}_2 \setminus n_{\mu_{i}} n_{x_2},h^{\rm C'}_1h^{\rm C'}_2 \setminus n'_{\mu_{i}} n_{x'_2}} \notag \\
&+\sum_{i=1,2}\left([K]_{n_{z_{i}},n'_{z_i}}-\frac{V_0 \lambda^2}{4} [\bar{D}^{\rm 1d}(\tilde{\alpha}_p)]_{n_{z_{i}},n'_{z_i}}\right) \notag \\
&\times [D^{\rm 1d}(\tilde{\alpha})]_{n_{x_2},n'_{x_2}}\delta_{h^{\rm C}_1h^{\rm C}_2 \setminus n_{z_{i}} n_{x_2},h^{\rm C'}_1h^{\rm C'}_2 \setminus n'_{z_{i}} n_{x'_2}},
\end{align}
where $\bar{D}^{\rm 1d}(\tilde{\alpha}_p)=D^{\rm 1d}(\tilde{\alpha}_p)+D^{\rm 1d}(\tilde{-\alpha}_p)$ and $\delta_{h^{\rm C}_1h^{\rm C}_2 \setminus n_{\mu_{i}} n_{x_2},h^{\rm C'}_1h^{\rm C'}_2 \setminus n'_{\mu_{i}} n_{x'_2}}$ denote the Kronecker delta over all indices except $n_{\mu_{i}}$ and $ n_{x_2}$. The one dimensional kinetic matrix is given analytically by
\begin{align}
[K]_{n,n'}&=\langle n|\hat{p}^2|n'\rangle/2m =\frac{\hbar \omega_{\rm ho}}{4}\left[(2n'+1)\delta_{nn'} \right. \notag \\
&-\left.\sqrt{(n'+1)(n'+2)}\delta_{n,n'+2}-\sqrt{n'(n'-1)}\delta_{n,n'-2}\right]
\end{align}

Similarly, $T'$ in $x$-direction is given by
\begin{align}
T'&=\langle\psi_{12}\psi_3|D^{\dagger}_3(\alpha_x)\sum_{i=1}^3H_{\rm m}(\bm{r}_i)D_2(\alpha_x)D_3(\alpha_x)|\psi_1\psi_{23}\rangle \notag \\
&=\langle\psi_{12}\psi_3|\sum_{i=1}^3H_{\rm m}(\bm{r}_i)D_2(\alpha_x)|\psi_1\psi_{23}\rangle \notag \\
&=\sum_{H_{12}h_{12}}\sum_{H'_{23}h'_{23}}\sum_{h_1h_2h_3}\sum_{h'_1h'_2h'_3}\sum_{h^{\rm C}_1h^{\rm C}_2h^{\rm C}_3}\sum_{h^{\rm C'}_1h^{\rm C'}_2h^{\rm C'}_3} \notag \\
&\times c^*_{H_{12}}c^*_{h_{12}}c^*_3c_{h_1}c_{h_{23}}c_{H_{23}}[U^{-1}]_{H_{12}h_{12}}^{h_{1}h_{2}}[U]^{H'_{23}h'_{23}}_{h'_{2}h'_{3}}\notag \\
&\times[C^{-1}]_{h_1}^{h^{\rm C}_1}[C^{-1}]_{h_2}^{h^{\rm C}_2}  [C^{-1}]_{h_3}^{h^{\rm C}_3}[C]_{h^{\rm C'}_1}^{h'_1}[C]_{h^{\rm C'}_2}^{h'_2}[C]_{h^{\rm C'}_3}^{h'_3} \notag \\
&\times\langle h^{\rm C}_1|\langle h^{\rm C}_2|\langle h^{\rm C}_3|\sum_{i=1}^3H_{\rm m}(\bm{r}_i)D_2(\alpha_x)|h^{\rm C'}_1\rangle|h^{\rm C'}_2\rangle|h^{\rm C'}_3\rangle
\end{align}
with 
\begin{align}
&\langle h^{\rm C}_1|\langle h^{\rm C}_2|\langle h^{\rm C}_3|\sum_{i=1}^3H_{\rm m}(\bm{r}_i)D_2(\alpha_x)|h^{\rm C'}_1\rangle|h^{\rm C'}_2\rangle|h^{\rm C'}_3\rangle \notag \\
&=\left(\sum_{n''_{x_2}}[K]_{n_{x_2},n''_{x_2}}[D^{\rm 1d}(\tilde{\alpha})]_{n''_{x_2},n'_{x_2}} \right. \notag \\
&+ (3+\frac{3\lambda^2}{2})V_0*[D^{\rm 1d}(\tilde{\alpha})]_{n_{x_2},n'_{x_2}}\notag \\
&-\left. \frac{V_0}{4}\sum_{n''_{x_2}}[\bar{D}^{\rm 1d}(\tilde{\alpha}_p)]_{n_{x_{2}},n''_{x_2}}[D^{\rm 1d}(\tilde{\alpha})]_{n''_{x_2},n'_{x_2}}\right) \notag \\
&\times \delta_{h^{\rm C}_1h^{\rm C}_2 \setminus n_{x_2},h^{\rm C'}_1h^{\rm C'}_2 \setminus n_{x'_2}} \notag \\
&+\sum_{\mu=x,y,i=1,2 \setminus x_2} \left([K]_{n_{\mu_{i}},n'_{\mu_i}}-\frac{V_0}{4}[\bar{D}^{\rm 1d}(\tilde{\alpha}_p)]_{n_{\mu_{i}},n'_{\mu_i}}\right) \notag \\
&\times [D^{\rm 1d}(\tilde{\alpha})]_{n_{x_2},n'_{x_2}}\delta_{h^{\rm C}_1h^{\rm C}_2 \setminus n_{\mu_{i}} n_{x_2},h^{\rm C'}_1h^{\rm C'}_2 \setminus n'_{\mu_{i}} n_{x'_2}} \notag \\
&+\sum_{i=1,2}\left([K]_{n_{z_{i}},n'_{z_i}}-\frac{V_0 \lambda^2}{4} [\bar{D}^{\rm 1d}(\tilde{\alpha}_p)]_{n_{z_{i}},n'_{z_i}}\right) \notag \\
&\times [D^{\rm 1d}(\tilde{\alpha})]_{n_{x_2},n'_{x_2}}\delta_{h^{\rm C}_1h^{\rm C}_2 \setminus n_{z_{i}} n_{x_2},h^{\rm C'}_1h^{\rm C'}_2 \setminus n'_{z_{i}} n_{x'_2}},
\end{align}
\subsection{Alternative approach for $T$}
 Alternatively, the tunneling parameter $T$ can be expressed as
\begin{equation} \label{eq:Ta}
T=\int dr d\theta r^2 \sin \theta \psi^{r}(r,\theta) f(r,\theta)
\end{equation}
where the function $f(r,\theta)$ is obtained by analytically integrating over the center-of-mass coordinate $\bm{R}$ and the azimuthal angle $\varphi$ of the relative motion:
\begin{align}
 f(r,\theta)&\equiv\int d \bm{R}_{12} \frac{1}{\sqrt{2 \pi}}\int_{0}^{\rm 2\pi} d \varphi \psi^{R}(\bm{R}_{12}) \notag \\ 
&\times \sum_{i=1,2}H_{\rm m}(\bm{r}_i)\psi_1(\bm{r}_1)\psi_2(\bm{r}_2-\bm{d}).
\end{align}
Here, $d$ describes the displacement and $\psi_i(\bm{r}_i)$ and $\psi^{R}(\bm{R}_{12})$ denote the ground free harmonic oscillator state of a single monomer, and of a dimer in the center of mass degree of freedom, respectively. 

In the case of tunneling in the $x$ direction $\bm d =a_{\rm latt} \hat{e}_x$, we get
\begin{align}
&f_x(r,\theta) \notag \\
&=\sqrt{2\pi} A_x e^{-\frac{r^2(\lambda \cos{\theta}^2+\sin{\theta}^2)}{4a_{\rm ho}^2}} \notag \\
&\times \left\{F_x \cdot I_{0}\left( \frac{r \sin{\theta}\sqrt{a_{\rm latt}^2-4 a_{\rm ho}^4 \pi^2/a_{\rm latt}^2}}{2 a_{\rm ho}^2}\right) \right. \notag  \\
&+ \left. C_x r \sin{\theta}\cdot I_{1} \left(\frac{a_{\rm latt}r\sin{\theta}}{2 a_{\rm ho}^2}\right)+I_{0}\left(\frac{a_{\rm latt}r\sin{\theta}}{2 a_{\rm ho}^2}\right) \right. \notag \\
&\times \left. \left[ B_x+G_x \cos\left(\frac{\pi r \cos(\theta)}{a_{\rm latt}}\right)+D_x r^2(\lambda^2\cos{\theta}^2+\sin{\theta}^2) \right]
\right\}
\end{align}
with
\begin{align}
A_x&=\frac{\lambda^{1/4}}{8 a_{\rm ho}^{11/2}(2 \pi)^{3/4}}e^{-\frac{3a_{\rm latt}^2}{8 a_{\rm ho}^2}-\frac{a_{\rm ho}^2(1+\lambda)\pi^2}{2 a_{\rm latt}^2\lambda}},
\end{align}
\begin{align}
B_x&=e^{\frac{a_{\rm ho}^2(1+\lambda)\pi^2}{2 a_{\rm latt}^2 \lambda}} \notag \\
&\times \left[-\frac{\hbar^2 5 a_{\rm latt}^2}{2m}+\frac{\hbar^2 12 a_{\rm ho}^2 (2+\lambda)}{2m}+8 a_{\rm ho}^4(2+\lambda^2)V_0\right],
\end{align}
\begin{align}
C_x=\frac{\hbar^2 8 a_{\rm latt}}{2m}e^{\frac{a_{\rm ho}^2(1+\lambda)\pi^2}{2 a_{\rm latt}^2 \lambda}},D_x=-\frac{C_x}{2a_{\rm latt}},
\end{align}
\begin{align}
F_x=-8V_0a_{\rm ho}^4 e^{\frac{a_{\rm ho}^2 \pi^2}{2 a_{\rm latt}^2 \lambda}},G_x=-8V_0a_{\rm ho}^4 \lambda^2 e^{\frac{a_{\rm ho}^2 \pi^2}{2 a_{\rm latt}^2}}.
\end{align}

\begin{figure}
    \centering
    \includegraphics[width=1.0\linewidth]{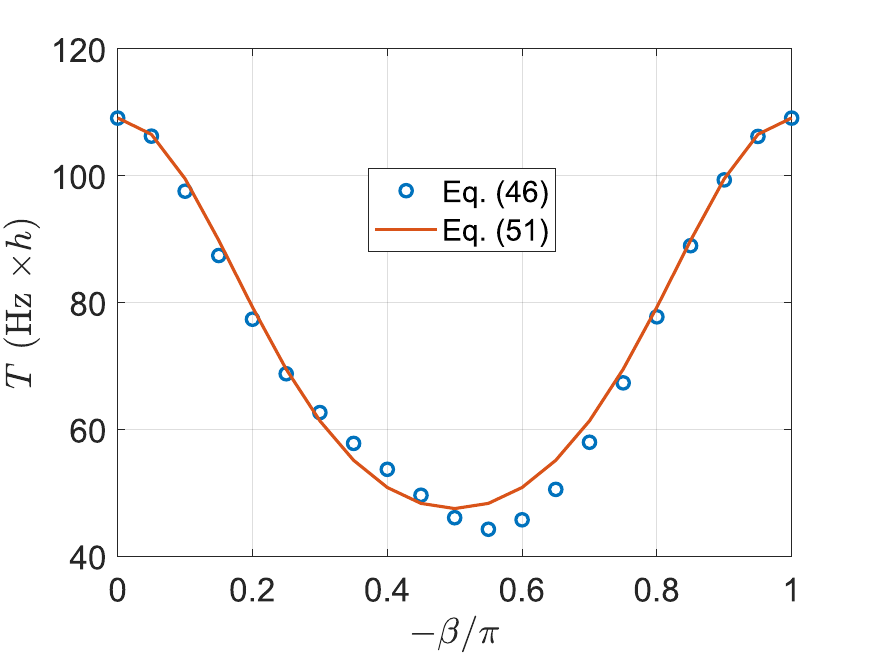}
    \caption{Comparison of $T$ obtained from harmonic oscillator bracket approach of Eq. \ref{eq:Th} (symbols) to that obtained from numerical integration according to Eq. \ref{eq:Ta}. The lattice parameters are chosen as $a_{\rm latt}=550$ nm, $s=5.1$ with $V_{0,z}=4V_{0,xy}$.}
    \label{fig:Tcom}
\end{figure}
Figure~\ref{fig:Tcom} shows that the values of $T$ obtained from the two approaches are in good agreement. Near $\beta=-\pi/2$, however, small numerical deviations cause the results obtained from Eq.~\ref{eq:Th} in the previous section to exhibit a slight asymmetry. In contrast, the approach introduced in this section provides improved numerical performance and yields results that are more symmetric about $\beta=-\pi/2$. Although extending this formalism to the three-monomer case required for evaluating $T'$ is not straightforward, the present calculations provide an important benchmark for validating the more general numerical method presented in the previous section.
\subsection{Dimer-monomer and dimer-dimer off-site interactions under rotation}
\begin{figure}
    \centering
    \includegraphics[width=1.0\linewidth]{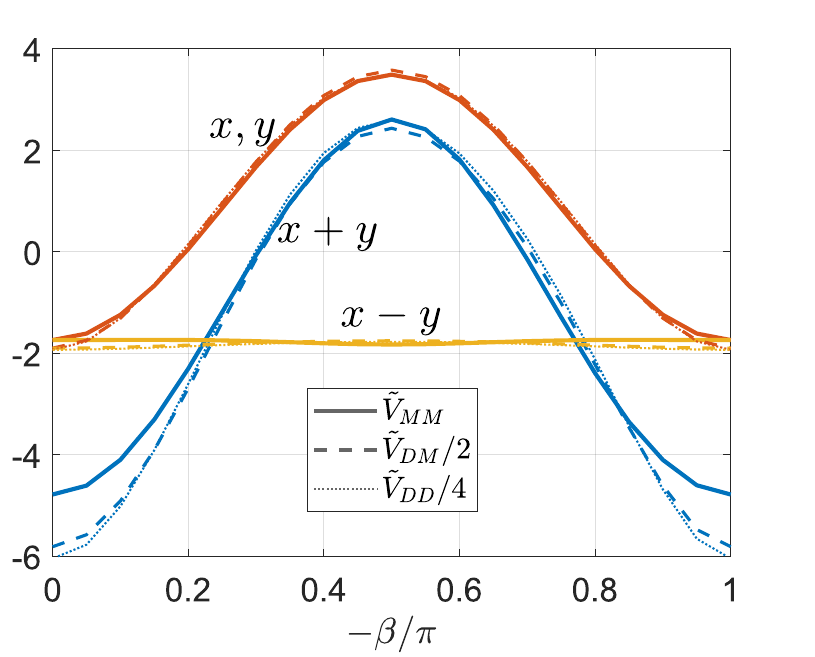}
    \caption{ Comparison of dimer-dimer and dimer-monomer to monomer-monomer interactions when rotating the microwave field frame ($\alpha=\pi/4$, $\beta$, $\gamma=-\pi/4$) relative to the lattice frame. In all cases nearest-neighbor ($x,y$) and diagonal next-nearest-neighbor ($x \pm y$) interactions are considered.}
    \label{fig:Vrot}
\end{figure}
We find that the dimer--monomer and dimer--dimer off-site interactions, $V_{DM}^{\sigma}$ and $V_{DD}^{\sigma}$, display essentially the same dependence on the rotation angle $\beta$ as the monomer--monomer interaction $V_{MM}^{\sigma}$ for all displacement configurations $\sigma = x$, $y$, and $x\pm y$, as illustrated in Fig.~\ref{fig:Vrot}. After scaling by the number of off-site monomer pairs, i.e., 2 for $V_{DM}^{\sigma}$ and 4 for $V_{DD}^{\sigma}$, the resulting quantities $V_{DM}^{\sigma}/2$ and $V_{DD}^{\sigma}/4$ closely match $V_{MM}^{\sigma}$ over the entire range of $\beta$. Discernible deviations occur only for the $\sigma = x$ and $y$ configurations when $\beta < -0.8\pi$ or $\beta > -0.2\pi$.

%

\end{document}